\documentclass[prd,draft,eqsecnum,showpacs,aps,nofootinbib,amsmath]{revtex4}

\newcommand{\de}{\partial}

\newcommand{\al}{\alpha}
\newcommand{\bt}{\beta}

\newcommand{\Gam}{\Gamma}
\newcommand{\del}{\delta}
\newcommand{\Del}{\Delta}

\newcommand{\zt}{\zeta}

\newcommand{\lam}{\lambda}
\newcommand{\Lam}{\Lambda}
\newcommand{\sig}{\sigma}
\newcommand{\vphi}{\varphi}
\newcommand{\om}{\omega}
\newcommand{\Om}{\Omega}

\newcommand{\0}[1]{{#1^{(0)}}}
\newcommand{\1}[1]{{#1^{(1)}}}
\newcommand{\2}[1]{{#1^{(2)}}}
\newcommand{\3}[1]{{#1^{(3)}}}

\renewcommand{\O}{{\mathcal O}}
\newcommand{\E}{{\mathcal E}}

\newcommand{\x}{\mathbf{x}}
\newcommand{\p}{\mathbf{p}}
\renewcommand{\d}{\mathbf{\hat{d}}}
\newcommand{\e}{\mathbf{\hat{e}}}

\renewcommand{\H}{\mathcal{H}}
\newcommand{\nab}{\nabla}
\newcommand{\Q}{\mathcal{Q}}
\renewcommand{\t}{\tilde}

\begin{document}

\title{CMB temperature anisotropies from third order gravitational perturbations}

\author{Guido D'Amico}
\email{damico@sissa.it}
\affiliation{SISSA/ISAS, via Beirut 2-4, 34014, Trieste, Italy}

\author{Nicola Bartolo}
\email{nicola.bartolo@pd.infn.it}
\affiliation{Dipartimento di Fisica ``G.\ Galilei'', Universit\`{a} di Padova \\
and INFN - Sezione di Padova, via Marzolo 8, I-35131 Padova, Italy}

\author{Sabino Matarrese}
\email{sabino.matarrese@pd.infn.it}
\affiliation{Dipartimento di Fisica ``G.\ Galilei'', Universit\`{a} di Padova \\
INFN - Sezione di Padova, via Marzolo 8, I-35131 Padova, Italy}

\author{Antonio Riotto}
\email{antonio.riotto@pd.infn.it}
\affiliation{INFN, Sezione di Padova, Via Marzolo, 8 - I-35131 Padua - Italy, \\
and D\'epartement de Physique Th\'eorique, Universit\'e de Gen\`eve, 24 Quai Ansermet, Gen\`eve, Switzerland}

\begin{abstract}
In this paper we present a complete computation of the Cosmic Microwave Background (CMB)
anisotropies up to third order from gravitational perturbations accounting for scalar,
vector and tensor perturbations. We then specify our results to the large scale
limit, providing the evolution of the gravitational potentials in a flat universe
filled with matter and cosmological constant which characterizes the
Integrated Sachs-Wolfe effect. As a byproduct in the large scale approximation
we are able to give non-perturbative solutions for the photon geodesic equations.
Our results are the first step to provide a complete theoretical
prediction for cubic non-linearities which are particularly relevant 
for characterizing the level of non-Gaussianity in the CMB through the detection of the
four-point angular connected correlation function (trispectrum). For this
purpose we also allow for generic initial conditions due to primordial
non-Gaussianity.

\end{abstract}

\pacs{98.80.Cq \hfill DFPD--A/07/14}
\maketitle

\section{Introduction}
The three year data set of the Wilkinson Anisotropy Probe (WMAP) on Cosmic Microwave Background (CMB) temperature anisotropies 
has offered a wealth of information about the evolution of the universe of unprecedented accuracy~\cite{kwmap3} and prepared the way for even more 
ambitious future missions such as those of the {\it Planck} satellite~\cite{Planck} and CMB polarization observations. Such precise measurements 
have been accompanied by increasing  theoretical efforts in predicting the signatures on the CMB from various cosmological scenarios. In particular 
a lot of attention in the last years has been dedicated to the statistical properties of the CMB beyond the power spectrum, 
in search for possible non-Gaussian signatures~\cite{review}. 
Non-Gaussianity of the CMB owes its importance to the possibility of unveiling crucial aspects of the
physics of both the early and the late universe, which would be unreachable using only the CMB power spectra information. Different mechanisms for the 
generation of the cosmological perturbations predict different amplitudes and shapes of primordial non-Gaussianity, thus a positive detection of NG or 
an upper limit on its amplitude is powerful in discriminating among the various competing scenarios which would be indistinguishable otherwise~\cite{review}.
An illuminating example is given by the standard  models of inflation which typically predict a very low content of non-Gaussianity~\cite{ABMR,maldacena} 
and as such they could be completely ruled out if a positive detection will be achieved.
CMB non-Gaussianity can also have a non-primordial origin due to secondary anisotropies
which arise when the CMB photons leave the last scattering surface and cross the large-scale structure of the Universe.
These include both secondary scatterings
(such as the thermal and kinetic Sunyaev-Zel' dovich effects, produced by the thermal and bulk motions of electrons in clusters,
and the Ostriker-Vishniac effect, due to bulk motions modulated by linear density perturbations) and gravitational secondaries such as the gravitational 
lensing and the non-linear Integrated Sachs-Wolfe effect or, on smaller scales, the Rees-Sciama effect due to the non-linear evolution of the potentials.   
Such non-linearities can give important and new information about the dark matter and the dark energy content of the universe~\cite{MGLM,Bern,
GS1,GS2,VerdeSDE,GBP,ks,fullT}. Statistics like the bispectrum (the Fourier transform of the three-point correlation function) and the trispectrum 
(four-point correlation) can then be used to assess the level of primordial non-Gaussianity on various cosmological scales 
and to discriminate it from the ones induced by the secondary anisotropies and systematic effects.

Up to now most of the attention has focused on the three-point statistics both for the primordial non-Gaussianity and for the analysis of the angular 
bispectrum of the CMB temperature and polarization anisotropies~\cite{review,ks,kwmap,babichpol,Liguorietal,Liguorietal2}.
On the other hand, the four-point correlation function could display interesting features. From the observational point of view it has been argued that it 
could be even more sensible to primordial non-Gaussianity than the bispectrum for very small angular scales in the next CMB 
experiments~\cite{OkamotoHu,KogoKomatsu}.
>From the theoretical point of view the four-point connected correlation function for the large scale CMB anisotropies has been 
computed in Ref.~\cite{Bartolo:2005fp} giving the theoretical predictions for the quadratic and cubic non-linearities 
(which include generic non-Gaussian initial conditions), while Refs.~\cite{SL1,SL2} have computed 
the trispectrum from single-field and uncoupled multiple fields in slow-roll inflation showing that it is of the order of the slow-roll parameters. 
Interestingly in some cases, such as in some configurations of the curvaton scenarios~\cite{SasakiVW,BSW}, 
the main source of a non-Gaussian signal can come from the four-point correlations functions, 
the three-point correlation function being suppressed (see also the phenomenological model discussed in Ref.~\cite{kwmap3} and Ref.~\cite{HuangShiu}).  

However most of these computations deal with the trispectrum of the uniform density curvature perturbation $\zeta$ 
on large scales within a given inflationary model, but this is  
not the physical quantity that is observed. The same is true of course for the bispectrum of the curvature perturbation. 
But, if for the case of the curvature perturbation bispectrum one is sure 
that the information about the primordial non-Gaussianity in the final observable quantity (the CMB anisotropies) is just the one obtained by evolving linearly 
the curvature bispectrum, the same is not true any more for the trispectrum. Let us consider the standard way to characterize the level of non-Gaussianity 
in the 
gravitational potential by expanding it as 
\begin{equation}
\label{fg}
\Phi=\Phi_L+f_{\rm NL}(\Phi^2_L-\langle \Phi^2_L\rangle)+g_{\rm NL} \Phi_L^3\, ,
\end{equation}       
where $\Phi_L$ is the linear first-order Gaussian part, 
$f_{\rm NL}$ and $g_{\rm NL}$ are the parameters which measure the quadratic and cubic non-linearities. In fact the non-linearity parameters might have 
in general 
a non trivial scale dependence. At linear order 
the curvature perturbation is $\zeta^{(1)}=5 \Phi_L/3$ during the matter dominated epoch, and, for example, the Sachs-Wolfe effect tells us 
that $\Delta T/T=-\Phi_L/3$. 
The standard approach followed in the literature is to extend both of these two relations also at higher-order, so that for example for quadratic 
non-Gaussianity
one writes $\zeta=5 \Phi/3$ using the expansion~(\ref{fg}) up to second-order. Evolving linearly the perturbations in such a way just accounts 
for the primordial content of non-Gaussianity and is justified only when one assumes that the primordial level of non-Gaussianity 
is much larger of the second-order corrections (in the form of first-order squared perturbations) which arise both in the relation between the curvature 
perturbation 
and the gravitational potential and when computing the CMB anisotropies. One is guaranteed that such primordial 
contribution in the final CMB anisotropies will appear as computed in this way because it is already an intrinsically second-order non-linearity and 
all the transfer 
functions will therefore be the same as at linear order. For the CMB bispectrum this is also all what is necessary to account for the primordial 
non-Gaussianity. 
For the trispectrum the situation is different, however. Evolving linearly the primordial cubic 
non-Gaussianity in Eq.~(\ref{fg}) would just lead to $\zeta^{(3)}/6=5g_{\rm NL}/3 \Phi_L^{3}$ 
(splitting $\zeta=\zeta^{(1)}+\zeta^{(2)}/2+...$ into a first- and higher order parts). Since the curvature perturbation remains constant on superhorizon 
scales 
(for adiabatic perturbations) it is a useful quantity to keep track of the primordial non-Gaussianity. Therefore, suppose to parametrize the primordial 
non-Gaussianity as  
\begin{equation}
\zeta=\zeta^{(1)}+ (a_{\rm NL}-1) \left( \1{\zt} \right)^2 +(b_{\rm NL}-1) \left( \1{\zt} \right)^3\, ,
\end{equation} 
where the two non-linearity parameters depend on the physics of a given scenario for the generation of the perturbations (for example, for standard 
single-field models of inflation $a_{\rm NL}=1$ and $b_{\rm NL}=1$, plus tiny corrections proportional to the slow-roll parameters). Thus the relation 
$\zeta^{(3)}/6=5g_{\rm NL}/3 \Phi_L^{3}$ is equivalent to
\begin{equation}
\label{gr}
g_{\rm NL}=\frac{25}{9} (b_{\rm NL}-1)\, .
\end{equation} 
However, this does not catch at all the whole information about the primordial non-Gaussianity that is contained in the CMB anisotropies up to third order. 
This is due to two reasons: first, as a source for the evolution of the gravitational potentials at third-order now there are also the gravitational 
potentials at second-order which inevitably contain primordial non-linearities proportional to the $f_{\rm NL}$ (or $a_{\rm NL}$) parameter;
and second, when computing the CMB anisotropies at third-order,
one still gets additional terms proportional to $f_{\rm NL} \Phi_L^3$.
This point is clear in the results we present in Sec.~\ref{3S} and~\ref{e3p}. 
See for example the first line of Eq.~(\ref{T3scalars}), or Eqs.~(\ref{source}) and~(\ref{A}).    
Another example can be found in Ref.~\cite{Bartolo:2005fp}, where a fully non-linear 
expression for the Sachs-Wolfe effect has been obtained for generic non-Gaussian initial conditions. The corresponding parameter $g_{\rm NL}$ for CMB 
anisotropies 
has been computed (see Eq.~(67) of that Reference) by accounting for all the possible dependence on the primordial non-Gaussianity finding an expression 
of the type 
\begin{equation}
\label{gT}
g_{\rm NL}= \frac{25}{9} (b_{\rm NL}-1)+A({\bf k}_1,{\bf k}_2,{\bf k}_3;a_{\rm NL})\, , 
\end{equation}
which, contrary to Eq.~(\ref{gr}), depends also on the quadratic non-linearity parameter
(here ${\bf k}_i$ are the perturbation wavenumbers in Fourier space). 
A similar expression to Eq.~(\ref{gT}) is given for the gravitational potentials at third-order in Eq.~(\ref{Phiba}). These examples show that even in the case 
of a ``local'' model for quadratic non-Gaussianity (i.e. a constant $f_{\rm NL}$ parameter), at higher orders such non-linearities are modulated by 
first-order perturbations generating a scale-dependent non-Gaussianity.

The best limits up to date on the $f_{\rm NL}$ parameter come from measurements of the CMB bispectrum on the WMAP data
giving $-36 \leq f_{\rm NL} \leq  100$~\cite{kwmap,kwmap3,Crem},
while there is at present no real bound on $g_{\rm NL}$. Given the increasing precision of future mesurements of CMB anisotropies, 
it is clear that it is of fundamental importance to provide accurate predictions for all the cubic non-linearities that enter in the 
evolution of the cosmological perturbations.
This is not only mandatory to be able to evaluate the trispectrum of CMB anisotropies,
but becomes crucial if one wants to account for the precise dependence of the trispectrum on the primordial non-Gaussianity.
Spurred by these considerations, in this paper we will focus on the CMB anisotropies up to 
third-order from gravitational perturbations due to the redshift the photons suffer when they travel from the last scattering surface to the observer. 
The computation will be performed following  Refs.~\cite{Pyne:1993np,Pyne:1995bs,Mollerach:1997up}
by perturbing at the desired order the photon geodesic 
accounting for scalar, vector and tensor perturbations.
However we will also propose an alternative method to solve the photon geodesic equation (for scalar perturbations), which consists in 
a non-perturbative computation. In the last part of the paper we focus on the large-scale approximation and on scalar perturbations for a Universe filled 
with non-relativistic matter and a non-vanishing cosmological constant.
In fact one of the goals is to characterize the evolution of the gravitational potentials on large scales which is 
responsible for the (late) Integrated Sachs-Wolfe effect, thus completing the analysis for the large scale CMB anisotropies at third-order already started in 
Ref.~\cite{Bartolo:2005fp}.
Of course it is out of the goals of this paper to deal with the anisotropies generated from the dynamics at recombination 
and due to the scattering terms~\footnote{A complete computation of the CMB anisotropies due to the non-linear dynamics
taking place at the last scattering epoch has been performed in Refs.~\cite{CMB2first,CMB2second},
where the full system of the Boltzmann equations up to second-order for photons, baryons and cold dark matter have been 
presented togheter with analytical solutions in the tight coupling approximation.}.

\section{Temperature anisotropies}
\label{2S}
We are interested in the pattern of fluctuations of the CMB temperature as measured by an observer
in a perturbed flat Friedmann-Robertson-Walker spacetime.
 
The line element can be written as 
\begin{equation}
 \label{metric}
 ds^2 = a^2(\eta) g_{\mu \nu} dx^\mu dx^\nu = a^2(\eta) \left[ \0{g_{\mu \nu}} + \1{g_{\mu \nu}} + \2{g_{\mu \nu}} + \3{g_{\mu \nu}} + \dots \right]
 	dx^\mu dx^\nu \, ,
\end{equation}
where $a(\eta)$ is the scale factor in conformal time $d\eta=dt/a(t)$, $\0{g_{\mu \nu}}$ is the background Minkowski metric,
and $g^{(r)}_{\mu \nu}$ is the $r$-th order metric perturbation.

Photons travel along null geodesics $x^\mu(\lam)$,
where $\lam$ is an affine parameter in the conformal metric $g_{\mu \nu}$.
The photon path connects the point of observation, with coordinates $x^\mu_\O = (\eta_\O, \x_\O)$,
to the hypersurface of emission, defined as the spacelike hypersurface of constant conformal time $\eta_\E$.
The actual last scattering surface is the intersection of the observer's past light cone with this hypersurface.
We assume that on the constant $\eta_\E$ hypersurface every point $\p$ emits thermal radiation,
characterized by a temperature $T(\p,\d)$ which depends on the point of emission $\p$
and on the direction of emission, described by the vector $\d$ normalized to unity in the conformal background metric.
The different photon paths are specified by the direction from which they arrive at $\O$,
described by a vector $\e$, normalized to unity in the conformal background metric;
this unit vector can be thought as the direction toward which the observer is pointing an antenna.
The initial conditions $x^\mu_\O$, $\e$ determine the point $\p$ and direction $\d$ of emission.

During their travel from the last scattering surface to the observer the CMB photons suffer a redshift determined by the ratio of the emitted 
frequency $\omega_{\E}$ and the observed one $\omega_{\mathcal O}$. For a blackbody spectrum, the ratio $\om / T$ is constant along the photon path,
and the temperature measured by an observer is given by
\begin{equation}
 \label{blackbody}
 T_\O (\x_\O, \e) = \frac{\om_\O}{\om_\E} T_\E (\p, \d) \, .
\end{equation}
The expression for the frequency is
\begin{equation}
 \label{frequency}
 \om = - g_{\mu \nu} u^\mu k^\nu \, ,
\end{equation}
where $u^\mu$ is the four-velocity of the observer or emitter, normalized to $a^2 g_{\mu \nu} u^\mu u^\nu = -1$,
and $k^\nu(\lam) = \frac{d x^\nu}{d \lam}$ is the photon wavevector, tangent to the null geodesic $x^\nu(\lam)$.

Given the initial conditions $x^\mu_\O, \e, \om_\O$, we need to compute the quantities $\p, \d, \om_\E$,
propagating the photons back from the observation to the emission surface.
These will depend on the photon path and the associated wavevector,
which we expand in series of the metric perturbations $g_{\mu \nu}^{(r)}$ and their derivatives:
\begin{align}
 x^\mu(\lam) =& x^{(0)\mu}(\lam) + x^{(1)\mu}(\lam) + x^{(2)\mu}(\lam) + x^{(3)\mu}(\lam) + \dots \\
 k^\mu(\lam) =& k^{(0)\mu}(\lam) + k^{(1)\mu}(\lam) + k^{(2)\mu}(\lam) + k^{(3)\mu}(\lam) + \dots
\end{align}

We find it useful to write the perturbed (conformal) metric as 
\begin{equation}
\label{metricdef}
 g_{\mu \nu} dx^\mu dx^\nu = - e^{2 \Phi} d\eta^2 + 2 \om_i d\eta dx^i
 	+ \left( e^{-2 \Psi} \del_{ij} + \chi_{ij} \right) dx^i dx^j\, ,
\end{equation}
which is valid at any order in perturbation theory. The two gravitational potentials $\Phi$ and $\Psi$ correspond to scalar metric perturbations, 
$\omega_i$ includes a scalar and a vector perturbation, and $\chi_{ij}$ contains another scalar, as well as vector and tensor perturbations.    
To make contact with the usual perturbative calculations one expands every quantity as
$\Phi = \1{\Phi} + \frac{1}{2} \2{\Phi} + \frac{1}{6} \3{\Phi} + \dots$ and analogously for the others.
Notice that, unlike $\om^{(r)}_i$ and $\chi^{(r)}_{ij}$, the quantities $\Phi^{(r)}$ and $\Psi^{(r)}$ 
are not the usual $r$-th order scalars $\phi^{(r)}$, $\psi^{(r)}$ appearing in Refs.~\cite{MMB,review}.  
However, it is immediate to find the relation between these quantities. Up to third order 
$\1{\Phi} = \1{\phi}$, $\1{\Psi} = \1{\psi}$, $\2{\Phi} = \2{\phi} - 2 \phi_{(1)}^2$,
$\2{\Psi} = \2{\psi} + 2 \psi_{(1)}^2$, $\3{\Phi} = \3{\phi} - 6 \1{\phi} \2{\phi} + 8 \phi_{(1)}^3$,
$\3{\Psi} = \3{\psi} + 6 \1{\psi} \2{\psi} + 8 \psi_{(1)}^3$. The form of the metric~(\ref{metricdef}) greatly helps in the intermediate computations, 
however some results will be expressed in the variables $\phi^{(r)}$ and $\psi^{(r)}$ because they appear in a more compact form. Notice that in this 
section we will not choose a particular gauge so all the following expressions for the CMB anisotropies are valid in any gauge.

The four-velocity will be expanded as
\begin{equation}
 u^\mu = \frac{1}{a} \left( \del_0^\mu + v^{(1)\mu} + \frac{1}{2} v^{(2)\mu} + \frac{1}{6} v^{(3)\mu} + \dots \right)\, .
\end{equation}

The zero component of the velocity is fixed from the normalization condition; we find
\begin{align}
 v^0_{(1)} =& - \1{\phi} \, ; \\
 v^0_{(2)} =& - \2{\phi} + 3 \left( \1{\phi} \right)^2
  		+ 2 \1{\om_i} v^i_{(1)} + v^i_{(1)} \1{v_i} \, ; \\
 v^0_{(3)} =& - \3{\phi} + 3 \2{\om_i} v^i_{(1)} + 3 \1{\om_i} v^i_{(2)} + 9 \2{\phi} \1{\phi}
  	+ 3 \left( \1{\chi_{ij}} - 2 \1{\psi} \del_{ij} \right) v^i_{(1)} v^j_{(1)} \nonumber \\
  	&- 12 \1{\phi} \1{\om_i} v^i_{(1)} - 15 \left( \1{\phi} \right)^3 - 3 \1{\phi} v^i_{(1)} \1{v_i}
  	+ 3 v^i_{(2)} \1{v_i} \, .
\end{align}

In order to compute the observed temperature up to third order in perturbations,
we expand the frequency as
\begin{equation}
 \om = \0{\om} \left( 1 + \1{\tilde{\om}} + \2{\tilde{\om}} + \3{\tilde{\om}} + \dots \right) \, ,
\end{equation}
and the temperature at emission as
\begin{equation}
 T_\E(\p, \d) = \0{T_\E} \left( 1 + \tau(\p, \d) \right) \, ,
\end{equation}
where $\tau = \1{\tau} + \2{\tau} + \3{\tau} + \dots$ is the intrinsic temperature fluctuation at emission.
A calculation of this quantity up to third-order is beyond the goal of this paper, since 
we are interested in the additional effects of gravity along the photon path.
However on large scales (bigger than the horizon at recombination) it has been computed in a non-pertubative way in Ref.~\cite{Bartolo:2005fp}.
Its fully non-linear expression is very simple: $\tau=e^{-2\Phi/3}-1$. Moreover, on smaller scales, a complete treatment of the CMB anisotropies 
including the acoustic oscillations on the surface of last scattering up to second-order have been studied analytically 
in Ref.~\cite{CMB2first,CMB2second}, and a full numerical evaluation 
can be performed Ref.~\cite{numerico} using the set of Boltzmann equations provided in Ref.~\cite{CMB2first}.

Now, we have to take into account that we need to expand the point $\p$ and direction $\d$ at emission as
$\p = \0{\p} + \1{\p} + \2{\p} + \dots$ and $\d = \0{\d} + \1{\d} + \2{\d} + \dots$;
then we Taylor expand $\tau(\p, \d)$ around the background values, obtaining
\begin{equation}
\begin{split}
 \tau(\p, \d) =& \1{\tau} +
 	\left[ \2{\tau} + p_{(1)}^i \frac{\de \1{\tau}}{\de x^i} + d_{(1)}^i \frac{\de \1{\tau}}{\de d^i} \right]\\
 	&+ \left[ \3{\tau} + p_{(1)}^i \frac{\de \2{\tau}}{\de x^i} + d_{(1)}^i \frac{\de \2{\tau}}{\de d^i}
 		+ \frac{1}{2} p_{(1)}^i p_{(1)}^j \frac{\de^2 \1{\tau}}{\de x^i \, \de x^j}
 		+ p_{(1)}^i d_{(1)}^j \frac{\de^2 \1{\tau}}{\de x^i \, \de d^j} \right. \\
 		& \left. \quad + \frac{1}{2} d_{(1)}^i d_{(1)}^j \frac{\de^2 \1{\tau}}{\de d^i \, \de d^j}
 			+ p_{(2)}^i \frac{\de \1{\tau}}{\de x^i} + d_{(2)}^i \frac{\de \1{\tau}}{\de d^i} \right]
\end{split}
\end{equation}
where it is understood that all quantities are evaluated at $\0{\p}$, $\0{\d}$.
We find $p^{(1)i} = x^{(1)i} - k^{(0)i} x^{(1)0}$,
$p^{(2)i} = x^{(2)i} - k^{(0)i} x^{(2)0} + k^{(0)i} k^{(1)0} x^{(1)0} - k^{(1)i} x^{(1)0}$,
$d^{(1)i} = \frac{k^{(0)i} + k^{(1)i}}{\left|k^{(0)i} + k^{(1)i} \right|}
- \frac{k^{(0)i}}{\left|k^{(0)i}\right|}$,
$d^{(2)i} = \frac{k^{(0)i} + k^{(1)i} + k^{(2)i}}{\left|k^{(0)i} + k^{(1)i} + k^{(2)i} \right|}
- \frac{k^{(0)i} + k^{(1)i}}{\left|k^{(0)i} + k^{(1)i} \right|}$.
Notice that therefore the quantities $\hat{\bf d}^{(i)}$ are just the 
difference between the photon normalized wavevector up to the $i$-th order and that at the $(i-1)$-th order.

Performing all the expansions in Eq.\eqref{blackbody}
and dividing by the background temperature $\frac{a_\E}{a_\O} \0{T_\E}$, we find
\begin{equation}
 \frac{\del \1{T}}{T} = \1{\tilde{\om}}_\O - \1{\tilde{\om}}_\E + \1{\tau} \, ;
\end{equation}
\begin{equation}
 \frac{\del \2{T}}{T} = \2{\tilde{\om}}_\O - \2{\tilde{\om}}_\E
 	+ \2{\tau} + p_{(1)}^i \frac{\de \1{\tau}}{\de x^i} + d_{(1)}^i \frac{\de \1{\tau}}{\de d^i}
 	+ \left( \1{\tau} - \1{\tilde{\om}}_\E \right) \left( \1{\tilde{\om}}_\O - \1{\tilde{\om}}_\E \right) \, ;
\end{equation}
\begin{equation}
\begin{split}
 \frac{\del \3{T}}{T} =& \, \3{\tilde{\om}}_\O - \3{\tilde{\om}}_\E
 	+ \3{\tau} + p_{(1)}^i \frac{\de \2{\tau}}{\de x^i} + d_{(1)}^i \frac{\de \2{\tau}}{\de d^i}
 	+ \frac{1}{2} p_{(1)}^i p_{(1)}^j \frac{\de^2 \1{\tau}}{\de x^i \, \de x^j} \\
 	& + p_{(1)}^i d_{(1)}^j \frac{\de^2 \1{\tau}}{\de x^i \, \de d^j}
 	+ \frac{1}{2} d_{(1)}^i d_{(1)}^j \frac{\de^2 \1{\tau}}{\de d^i \, \de d^j}
 		+ p_{(2)}^i \frac{\de \1{\tau}}{\de x^i} + d_{(2)}^i \frac{\de \1{\tau}}{\de d^i} \\
 	& + \left( \2{\tau} + p_{(1)}^i \frac{\de \1{\tau}}{\de x^i}
 		+ d_{(1)}^i \frac{\de \1{\tau}}{\de d^i} - \2{\tilde{\om}}_\E \right)
 		\left( \1{\tilde{\om}}_\O - \1{\tilde{\om}}_\E \right) \\
 	& + \left( \1{\tau} - \1{\tilde{\om}}_\E \right)
 		\left( \2{\tilde{\om}}_\O - \2{\tilde{\om}}_\E + (\1{\tilde{\om}}_\E)^2
 		- \1{\tilde{\om}}_\E \1{\tilde{\om}}_\O \right) \, .
\end{split}
\end{equation}

The expansion of the frequency using Eq.(\ref{frequency}) yields:
\begin{equation}
 \1{\tilde{\om}_\E} = k^0_{1} + \1{\phi} + \1{\om_i} e^i + \1{v_i} e^i \, ;
\end{equation}

\begin{equation}
\begin{split}
 \2{\tilde{\om}_\E} =& k^0_{(2)} + \frac{1}{2} \2{\phi} + \frac{1}{2} \2{\om_i} e^i
  	 + \frac{1}{2} v^i_{(1)} \1{v_i} - \frac{1}{2} (\1{\phi})^2
  	 + \1{\phi} k^0_{(1)} - \1{\om_i} k^i_{(1)} - \1{v_i} k^i_{(1)} \\
  	 &- \1{\phi} \1{\om_i} e^i - 2 \1{\psi} \1{v_i} e^i
  	 + \1{\chi_{ij}} v^i_{(1)} e^j + \frac12 \2{v_i} e^i \\
  	 &+ \1{\Del \lam} \frac{d k^0_{(1)}}{d \lam}
  	 + p^i_{(1)} \left( \de_i \1{\phi} + \de_i \1{\om_j} e^j + \de_i \1{v_j} e^j \right) \, ;
\end{split}
\end{equation}

\begin{equation}
\begin{split}
 \3{\tilde{\om}_\E} =& k^0_{(3)} + \1{\Del \lam} \frac{d k^0_{(2)}}{d \lam}
  		+ \2{\Del \lam} \frac{d k^0_{(1)}}{d \lam}
  		+ \frac{1}{2} \left( \1{\Del \lam} \right)^2 \frac{d^2 k^0_{(1)}}{d \lam^2} \\
  &+ \frac16 \3{\phi} + \frac12 p^i_{(1)} \de_i \2{\phi}
  		+ p^i_{(2)} \de_i \1{\phi} + \frac12 p^i_{(1)} p^j_{(1)} \de_i \de_j \1{\phi} \\
  &+ e^i \left( \frac16 \3{v_i} + \frac12 p^j_{(1)} \de_j \2{v_i}
  		+ p^j_{(2)} \de_j \1{v_i}
  		+ \frac{1}{2} p^j_{(1)} p^k_{(1)} \de_j \de_k \1{v_i} \right) \\
  &- \frac12 \2{\phi} \1{\phi}
  	+ \frac12 \left( \1{\chi_{ij}} - 2 \1{\psi} \del_{ij} \right) v^i_{(1)} v^j_{(1)} \\
  &+ \frac12 \left( \1{\phi} \right)^3 + \frac12 v^i_{(1)} \2{v_i}
  	+ \frac12 \1{\phi} \1{v_i} v^i_{(1)}
  	- \1{\phi} p^i_{(1)} \de_i \1{\phi} \\
  &+ \frac12 k^0_{(1)} \left[ \2{\phi} + 2 p^i_{(1)} \de_i \1{\phi}
  	- \left( \1{\phi} \right)^2 + v^i_{(1)} \1{v_i} \right]
  	+ \1{\phi} \left( k^0_{(2)} + \1{\Del \lam} \frac{d k^0_{(1)}}{d \lam} \right) \\
  & + \frac16 \3{\om_i} e^i + \frac12 p^i_{(1)} \de_i \2{\om_j} e^j
  		+ p^i_{(2)} \de_i \1{\om_j} e^j
  		+ \frac{1}{2} p^i_{(1)} p^j_{(1)} \de_i \de_j \1{\om_k} e^k \\
  &- \left( k^i_{(1)} + e^i \1{\phi} \right) \left( \frac12 \2{\om_i} + p^j_{(1)} \de_j \1{\om_i} \right)
  	- \1{\om_i} \left( k^i_{(2)} + \1{\Del \lam} \frac{d k^i_{(1)}}{d \lam} \right) \\
  &+ \1{\om_i} \Bigg\{ \1{\phi} k^i_{(1)} + \frac12 e^i \Big[ - \2{\phi} - 2 p^j_{(1)} \de_j \1{\phi} 
  	+ 3 \left( \1{\phi} \right)^2 + 2 \1{\om_j} v^j_{(1)} + v^j_{(1)} \1{v_j} \Big] \Bigg\} \\
  & - \del_{ij} k^i_{(1)} \left( \frac12 v^j_{(2)} + p^k_{(1)} \de_k v^j_{(1)} \right)
  	 	- \del_{ij} v^i_{(1)} \left( k^j_{(2)} + \1{\Del \lam} \frac{d k^j_{(1)}}{d \lam} \right) \\
  &+ \left( \1{\chi_{ij}} - 2 \1{\psi} \del_{ij} \right)
  		\left( \frac12 v^i_{(2)} e^j + p^k_{(1)} \de_k v^i_{(1)} e^j - v^i_{(1)} k^j_{(1)} \right) \\
  &+ \left[ \left( \frac12 \2{\chi_{ij}} - \2{\psi} \del_{ij} \right)
  	 	+ p^k_{(1)} \de_k \left( \1{\chi_{ij}} - 2 \1{\psi} \del_{ij} \right) \right] v^i_{(1)} e^j \, .
\end{split}
\end{equation}
In these expressions, $\1{\Del \lam}$ and $\1{\Del \lam}+ \2{\Del \lam}$ are the differences
in affine parameter between the points where the first order (respectively, the second order)
and the background geodesics intersect the $\eta = \eta_\E$ hypersurface.
They are given by $\1{\Del \lam} = - x^{(1)0}$, $\2{\Del \lam} = - x^{(2)0} + k^{(1)0} x^{(1)0}$.

The next step is to obtain the null geodesics up to third order,
using the formalism set up in \cite{Pyne:1995bs,Pyne:1993np}. This will allow us to compute at the desired order 
the photon path $x^{\mu}(\lambda)$ and the associated wavevector $k^{\mu}(\lambda)$ 
appearing in the previous expressions for the perturbations of the  photon frequency. The $a$-th order geodesic equation can be recasted 
as the forced Jacobi equation~\cite{Pyne:1995bs,Pyne:1993np}
\begin{equation}
 \frac{d x^{(a)\mu}}{d \lam^2} + 2 \Gam^{(0)\mu}_{\al \bt} k^{(0)\al} k^{(a)\bt}
 + \de_\sig \Gam^{(0)\mu}_{\al \bt} k^{(0)\al} k^{(0)\bt} x^{(a)\sig} = f^{(a)\mu} \, .
\end{equation}

In a flat background, the solutions are
\begin{align}
\label{xk}
 x^{(a) \mu} =& \left(\lam - \lam_\O \right) k^{(a) \mu} (\lam_\O)
 	+ \int_{\lam_\O}^\lam (\lam - \tilde{\lam}) f^{(a)\mu} (\tilde{\lam}) d\tilde{\lam} \, , \\
 k^{(a) \mu} =& k^{(a) \mu} (\lam_\O) + \int_{\lam_\O}^\lam f^{(a)\mu} (\tilde{\lam}) d\tilde{\lam} \, .
\end{align}

The forcing terms, in a flat background, are given by
\begin{align}
\label{fterms}
 f^\mu_{(1)} =& - \Gam^{(1)\mu}_{\al \bt} k^{(0)\al} k^{(0)\bt} \, , \\
 f^\mu_{(2)} =& - 2 \Gam^{(1)\mu}_{\al \bt} k^{(0)\al} k^{(1)\bt}
 -  x^{(1)\sig} \de_\sig \Gam^{(1)\mu}_{\al \bt} k^{(0)\al} k^{(0)\bt}
 - \Gam^{(2)\mu}_{\al \bt} k^{(0)\al} k^{(0)\bt} \, , \\
\begin{split}
\label{fterms3}
 f^\mu_{(3)} =& - \Gam^{(1) \mu}_{\al \bt} \left( 2 k^{\al}_{(2)} k^{\bt}_{(0)} + k^{\al}_{(1)} k^{\bt}_{(1)} \right)
 	- 2 k^{\al}_{(1)} k^{\bt}_{(0)} \left( \Gam^{(2) \mu}_{\al \bt}
 	+ \de_\sig \Gam^{(1) \mu}_{\al \bt} x^\sig_{(1)} \right) \\
 	&- k^\al_{(0)} k^\bt_{(0)} \left( \Gam^{(3) \mu}_{\al \bt}
 	+ \de_\sig \Gam^{(1) \mu}_{\al \bt} x^\sig_{(2)}
 	+ \de_\sig \Gam^{(2) \mu}_{\al \bt} x^\sig_{(1)}
 	+ \frac12 \de_\sig \de_\tau \Gam^{(1) \mu}_{\al \bt} x^\sig_{(1)} x^\tau_{(1)} \right) \, .
\end{split}
\end{align}
 The Christoffel symbols are calculated in the Appendix,
where one can find their expressions for the conformal metric \eqref{metricdef}. Notice 
that we have  recasted them in a compact form without specifying the order of the perturbations, even tough they can be used just up 
to third-order (at least for the terms coming from the vector and tensor perturbations). The method that uses Eqs.~(\ref{fterms}) is an iterative method:
once one knows the geodesic equations of $r-th$ oder, he or she is able to determine those at the next-order through Eqs.~(\ref{xk}).
Thus, in order to have all the means to compute the photon geodesics up to third order in the following,
we report the expressions that we have computed for the wavevectors $k^{(\mu)}$ up to second-order :
\begin{equation}
\label{k10}
 k^0_{(1)} = - 2 \1{\Phi} - \1{\om_i} e^i + \int_{\lam_\O}^\lam d\tilde{\lam} \1{A}'(\tilde{\lam})
    = - 2 \1{\Phi} - \1{\om_i} e^i + I_1(\lam)\, ,
\end{equation}
\begin{equation}
\label{k1i}
 k^i_{(1)} = - 2 \1{\Psi} e^i - \om_{(1)}^i + \chi_{(1)}^{ij} e_j
    - \int_{\lam_\O}^\lam d\tilde{\lam} \de^i \1{A}(\tilde{\lam})
    = - 2 \1{\Psi} e^i - \om_{(1)}^i + \chi_{(1)}^{ij} e_j - I_1^i(\lam) \, ,
\end{equation}
for the first-order wavevectors, and 
\begin{equation}
\begin{split}
\label{k20}
 k^0_{(2)} =& - \2{\Phi} - 2 \1{\Phi}^2 - \frac12 \2{\om_i} e^i - 2 k^0_{(1)} \1{\Phi}
    - 2 x_{(1)}^\mu \de_\mu \1{\Phi} + \1{\om_i} k^i_{(1)} - x_{(1)}^\mu \de_\mu \1{\om_i} e^i \\
    &+ \int_{\lam_\O}^\lam d\tilde{\lam} \Bigg[ \frac12 \2{A}'
        + 2 \left( \1{\Phi} \1{\Phi}' - \1{\Psi} \1{\Psi}'\right)
        + \left( \1{\chi}'_{ij} e^j - \1{\om_i}' \right) \left( e^i k^0_{(1)} + k^i_{(1)} \right) \\
    &\qquad \qquad + 2 k_{(1)}^0 \1{A}' + 2 \1{\Psi}' \1{A} + x_{(1)}^\mu \de_\mu \1{A}' \Bigg]\, ,
\end{split}
\end{equation}
\begin{equation}
\begin{split}
\label{k2i}
 k^i_{(2)} =& - e^i \2{\Psi} + 2 e^i \1{\Psi}^2 - 2 e^i \1{x}^\mu \de_\mu \1{\Psi} + 2 \1{k}^i \1{\Psi} \\
    &+ \frac12 \chi^i_{(2)j} e^j + x^\mu \de_\mu \chi^i_{(1)j} e^j - \chi^i_{(1)j} \1{k}^j
    - \frac12 \om_{(2)}^i - x_{(1)}^\mu \de_\mu \om_{(1)}^i - k_{(1)}^0 \om_{(1)}^i \\
    &+ \int_{\lam_\O}^\lam d\tilde{\lam} \Bigg[- \frac12 \de_i \2{A} - x^\mu_{(1)} \de_\mu \de^i \1{A}
    - 2 \1{k}^0 \de^i \1{\Phi} + \de^i \left( \1{\Psi}^2 - \1{\Phi}^2 \right) \\
    &\qquad \qquad + 2 e_j \1{k}^j \de^i \1{\Psi} - e^j \1{k}^k \de^i \1{\chi}_{jk}
    + k_{(1)}^j \de^i \om^{(1)}_j - k_{(1)}^0 \de^i \1{\om_j} e^j \Bigg] \, ,
\end{split}
\end{equation}
at second-order, where 
\begin{equation}
 A^{(r)} \equiv \Phi^{(r)} + \Psi^{(r)} + \om^{(r)}_i e^i - \frac12 \chi^{(r)}_{ij} e^i e^j \, .
\end{equation}

As a non-trivial test for the correctness of the above expressions, we have explicitly verified that the first 
and second order wavevectors do satisfy the null vector condition $k^\mu k_\mu = 0$.
When expanded perturbatively, this reads at linear order
\begin{equation}
  k_{(1)}^0 + e_i k_{(1)}^i = - A^{(1)} \, ,
\end{equation}
and at second order
\begin{equation}
\begin{split}
 &k_{(2)}^0 + e_i k_{(2)}^i + \frac{1}{2} (k_{(1)}^0)^2 - \frac{1}{2} k_{(1)}^i k^{(1)}_i + 2 \Phi_{(1)} k_{(1)}^0 - \om^{(1)}_i k_{(1)}^i
 	+ \om^{(1)}_i e^i k_{(1)}^0 - 2 \Psi_{(1)} e_i k_{(1)}^i + \chi^{(1)}_{ij} e^i k_{(1)}^j \\
 	&+ \frac{1}{2} A_{(2)} + \Phi_{(1)}^2 - \Psi_{(1)}^2 + x_{(1)}^\mu \de_\mu A_{(1)}= 0 \, .
\end{split}
\end{equation}
Both equations are satisfied when one substitutes the expressions~\eqref{k10}-\eqref{k2i}.

\section{CMB anisotropies on large scales from scalar perturbations}
\label{3S}
%\subsection{Temperature anisotropies}

In this Section, we focus on scalar perturbations and we give the expressions for the CMB anisotropies on large scales. 
Two main results are achieved. First we are able to give, using a non-perturbative method, an integral solution for the photon 
geodesic equations which holds at any order in perturbation theory. Second, 
we solve for the first time the evolution of the third-order gravitational potentials in a universe filled with non-relativistic 
matter and a cosmological constant. The presence of a non-vanishing cosmological constant makes the gravitational potentials vary at late 
times producing on large scales an Integrated-Sachs-Wolfe effect which is the counterpart of what happens 
at linear order. Of course, due to the non linear evolution of the perturbations, other integrated terms will appear in the form of cubic 
corrections. The ISW constitutes the main effect for large scale CMB anisotropies together with the Sachs-Wolfe effect originating at the last 
scattering surface. In fact both of them keep memory of the initial primordial non-Gaussianity of the cosmological perturbations 
(see details in Sec.~\ref{IC}), and as such  our expressions can be of particular relevance when trying to pin down the primordial 
non-Gaussian content from  the higher-order statistics of the CMB anisotropies as the bispectrum or the trispectrum. 
Let us see in details these two results.

 In the limit of large scales, it is convenient to adopt a non-perturbative formalism
to get the temperature anisotropies and the Einstein equations.
We take the metric to be $ds^2 = a^2(\eta)[- e^{2 \Phi} d\eta^2 + e^{-2 \Psi} \del_{ij} dx^i dx^j]$
and for simplicity we take the points of emission and observation to be comoving
(in such a way we lose the Doppler effects, however they are important on scales smaller than those we are interested here). In writing this 
metric we have neglected vector and tensor perturbation modes. For the vector perturbations the reason is that we are interested in 
long-wavelength perturbations, while vector modes will contain gradient terms being produced as non-linear combination of scalar modes 
and thus they will be more important on small scales (linear vector modes are not generated in standard mechanism for comsological 
perturbations, as inflation). For example the results of Ref.~\cite{MHMpol} show clearly this for second-order perturbations. 
In order to study the CMB anisotropies from scalar perturbations in the large scale limit, 
the tensor contribution can be negleted, since on large scales it has been proven to remain constant and to give a negligible 
effect being of 
the order of (powers of) the slow-roll parameters during inflation~\cite{SalopekBond,maldacena}.  

We start from eq. \eqref{blackbody} and we write the frequency as
\begin{equation}
 \om = - g_{\mu \nu} u^\mu k^\nu = e^{2 \Phi} u^0 k^0 = e^{\Phi} k^0 \, ,
\end{equation}
where $u^0=e^{-\Phi}$ is given by the normalization condition. The observed temperature is given by
\begin{equation}
 T_\O (\x,\e) = \frac{a_\E}{a_\O} e^{\Phi_\O-\Phi_\E} \frac{k^0_\O}{k^0_\E} T_\E (\p,\d) \, .
\end{equation}

We make another simplification: namely, we reabsorb the terms computed at the point of observation
into a redefinition of $T_\O$. 
This amounts to lose the monopole term, which however is unobservable.

At this point, we can perform the double expansion in perturbation orders and around the background geodesic
(i.e., the line of sight). We find
\begin{equation}
 \frac{\1{\del} T}{T} = \1{\tau} - \1{\Phi_\E} - k^{(1)0}_\E \, ;
\end{equation}

\begin{equation}
\label{T2scalars}
\begin{split}
 \frac{\2{\del} T}{T} =& \2{\tau} + p_{(1)}^i \frac{\de \1{\tau}}{\de x^i} + d_{(1)}^i \frac{\de \1{\tau}}{\de d^i}
 		- \frac12 \2{\Phi_\E} - p_{(1)}^i \frac{\de \1{\Phi_\E}}{\de x^i} + \frac{1}{2} (\1{\Phi_\E})^2 \\
 	& - k^{(2)0}_\E - \1{\Del \lam} \frac{d k^{(1)0}_\E}{d \lam} + (k^{(1)0}_\E)^2
 		- \1{\tau} \1{\Phi_\E} - \1{\tau} k^{(1)0}_\E + \1{\Phi_\E} k^{(1)0}_\E \, ;
\end{split}
\end{equation}

\begin{equation}
\label{T3scalars}
\begin{split}
 &\frac{\3{\del} T}{T} = \3{\tau} - k^{(3)0}_\E - \frac{1}{6} \3{\Phi_\E}
 	+ \frac{1}{2} \1{\Phi_\E} \2{\Phi_\E} - \frac{1}{6} \left( \1{\Phi_\E} \right)^3
 	+ 2 k^{(1)0}_\E k^{(2)0}_\E - (k^{(1)0}_\E)^3 - \2{\tau} \1{\Phi} \\
 	&+ \left[ k^{(2)0}_\E - (k^{(1)0}_\E)^2 \right] \left(\1{\Phi}- \1{\tau} \right)
 	+ \frac{1}{2} \1{\tau} \left[ - \2{\Phi_\E} + \left( \1{\Phi_\E} \right)^2 \right] \\
 	&+ k^{(1)0} \left[ - \2{\tau} + \frac{1}{2} \2{\Phi_\E} - \frac{1}{2} \left( \1{\Phi_\E} \right)^2
 		+ \1{\tau} \1{\Phi} \right] \\
 	&+ p_{(1)}^i \left[ \de_i \2{\tau} - \frac{1}{2} \de_i \2{\Phi} - \1{\tau} \de_i \1{\Phi}
 		- \1{\Phi} \de_i \1{\tau}
 		+ k^{(1)0} \left( \de_i \1{\Phi}- \de_i \1{\tau} \right)+ \1{\Phi} \de_i \1{\Phi} \right] \\
 	&+ d_{(1)}^i \left[ \frac{\de \2{\tau}}{\de d^i}
 		- \left( k^{(1)0} + \1{\Phi} \right) \frac{\de \1{\tau}}{\de d^i} \right]
 	+ \frac{1}{2} p_{(1)}^i p_{(1)}^j \left( \de_i \de_j \1{\tau} - \de_i \de_j \1{\Phi} \right) \\
 	&+ \frac{1}{2} d_{(1)}^i d_{(1)}^j \frac{\de^2 \1{\tau}}{\de d^i \, \de d^j}
 	+ p_{(2)}^i \left( \de_i \1{\tau} - \de_i \1{\Phi} \right)
 	+ d_{(2)}^i \frac{\de \1{\tau}}{\de d^i} + p_{(1)}^i d_{(1)}^j \frac{\de^2 \tau_{(1)}}{\de x^i \de d^j} \\
 	&+ \1{\Del \lam} \left[ - \frac{d k^{(2)0}_\E}{d \lam}
 		- \frac{1}{2} \1{\Del \lam} \frac{d^2 k^{(1)0}_\E}{d \lam^2}
 		+ \left( 2 k^{(1)0}_\E + \1{\Phi} - \1{\tau} \right) \frac{d k^{(1)0}_\E}{d \lam}\right]
 	- \2{\Del \lam} \frac{d k^{(1)0}_\E}{d \lam} \, .
\end{split}
\end{equation} 

In order to compute the photon wavevectors and the null geodesics, we will proceed in a different way than the usual standard perturbative method 
introduced in \cite{Pyne:1995bs,Pyne:1993np} which makes use of the Eqs.~(\ref{fterms})-(\ref{fterms3}). 
Let us instead start from a fully non-linear geodesic equation (time component) which can be written as
\begin{equation}
\label{geonp}
 \frac{d k^0}{d \lam} = - 2 \frac{d \Phi}{d \lam} k^0 + \left( \Phi' + \Psi' \right) \left(k^0\right)^2
\end{equation}
where we have used the null vector condition $g_{\mu \nu} k^\mu k^\nu = 0$. We can write the formal solution (formal since the potentials depend on the true geodesic)
as 
\begin{equation}
\label{formalsol}
 k^0(\lam) = e^{-2\Phi(\lam)} \left[1
 	- \int_{\lam_\O}^\lam e^{-2\Phi(x(\t{\lam}))}
 	\left( \Phi'(x(\t{\lam})) + \Psi'(x(\t{\lam})) \right) d\t{\lam} \right]^{-1} \, .
\end{equation}

The advantage of Eq.~\eqref{formalsol} is that it can be straightforwardly expanded to find
the perturbations to the background wavevector (and, by integration, to the background geodesic),
up to any desired order. To make contact with the standard perturbative results, in addition to the expansion in
perturbative orders we need to perform a Taylor expansion of the true geodesic around the background geodesic
for the potentials, in order to express all the quantities on the background geodesic.
For example, if we have the non-linear quantity $\al(x)$ computed on the true
geodesic, its expansion up to second order will be
$\al(x) = \al^{(1)} (x^{(0)}) + \de_\mu \al^{(1)} (x^{(0)})\, x^{(1)\mu} +
\al^{(2)} (x^{(0)})$, and similarly at higher-orders.

We find 
\begin{equation}
\label{k01}
 k^{(1)0}(\lam) = - 2 \1{\Phi} + I_1(\lam) \, ,
\end{equation}

\begin{equation}
\label{k02}
 k^{(2)0}(\lam) = - \2{\Phi} + 2 \left( \1{\Phi} \right)^2 - 2 \1{x}^\mu \de_\mu \1{\Phi}
 + I_2(\lam) - 2 \1{\Phi} I_1(\lam) + I_1(\lam)^2 \, ,
\end{equation}

\begin{equation}
\begin{split}
\label{k03}
 k^{(3)0}(\lam) =& - \frac{1}{3} \3{\Phi} - \1{x}^\mu \de_\mu \2{\Phi} - \1{x}^\mu \1{x}^\nu \de_\mu \de_\nu \1{\Phi}
 	- 2 \2{x}^\mu \de_\mu \1{\Phi} + \frac{8}{3} \left( \1{\Phi} \right)^3 \\
 &+ \left( 2 \1{\Phi} - I_1(\lam) \right)
 	\left[ \2{\Phi} - 2 \left( \1{\Phi} \right)^2 + 2 \1{x}^\mu \de_\mu \1{\Phi} \right] \\
 & + I_3(\lam) + 2 I_1(\lam) I_2(\lam) + I_1(\lam)^3 - 2 \1{\Phi} \left(I_2(\lam) + I_1(\lam)^2 \right) \, ,
\end{split}
\end{equation}
where
\begin{align}
\label{I1}
 I_1(\lam) =& \int_{\lam_\O}^\lam d\t{\lam} \1{A}'(\t{\lam}) \, , \\
\label{I2}
 I_2(\lam) =& \int_{\lam_\O}^\lam d\t{\lam} \left[ \frac{1}{2} \2{A}'(\t{\lam})
 + \1{x}^\mu \de_\mu \1{A}'(\t{\lam}) - 2 \1{\Phi} \1{A}'(\t{\lam}) \right] \, , \\
 \begin{split}
\label{I3}
 I_3(\lam) =& \int_{\lam_\O}^\lam d\t{\lam} \Big[ \frac{1}{6} \3{A}' + \frac{1}{2} \1{x}^\mu \de_\mu \2{A}'
 		+ \frac{1}{2} \1{x}^\mu \1{x}^\nu \de_\mu \de_\nu \1{A}' + \2{x}^\mu \de_\mu \1{A}' \\
 	&\qquad \quad - 2 \1{\Phi} \left( \frac{1}{2} \2{A}' + \1{x}^\mu \de_\mu \1{A}' \right)
 		+ \1{A}' \left( 2 \1{\Phi}^2 - \2{\Phi} - 2 \1{x}^\mu \de_\mu \1{\Phi} \right)\Big] \, ,
 \end{split} \\
\label{An}
 A^{(n)} =& \Phi^{(n)} + \Psi^{(n)} \, .
\end{align}
The above expressions are completely known once we find the geodesic $x^{\mu}(\lambda)$ (it is sufficient to compute them up to second-order, 
as it is clear by looking at Eq.(\ref{I3})). For the time component the 
integration of  $k^{(0)}(\lambda)$ yields      
\begin{equation}
 x^{(1)0}(\lam) = \int_{\lam_\O}^{\lam} d\t{\lam} \left[- 2 \1{\Phi}
 	+ \left(\lam - \t{\lam}\right) \1{A}' \right] \, ,
\end{equation}

\begin{equation}
\begin{split}
 x^{(2)0}(\lam) =& \int_{\lam_\O}^{\lam} d {\tilde \lambda}
 	\left[- \2{\Phi} + 2 \left( \1{\Phi} \right)^2 - 2 \1{x}^\mu \de_\mu \1{\Phi} - 2 \1{\Phi} I_1(\t{\lam}) + I_1(\t{\lam})^2 \right. \\
 &\qquad \quad + \left. \left(\lam - \t{\lam}\right) \left( \frac{1}{2} \2{A}'(\t{\lam}) + \1{x}^\mu \de_\mu \1{A}'(\t{\lam})
 	- 2 \1{\Phi} \1{A}'(\t{\lam}) \right) \right] \, .
\end{split}
\end{equation}
Notice that from Eq.(\ref{geonp}) to Eq.(\ref{I3}) we have never used the large scale approximation, so these expressions are general and 
we recover exactly the corresponding expressions for the second-order quantities already found in Refs.~\cite{Pyne:1993np,Mollerach:1997up}. 
On the other hand to compute the spatial components $x^i(\lambda)$ we do make use of the large scale approximation in the following way. 
We start from the non-perturbative spatial geodesic equation  
\begin{equation}
\label{gsnp}
	\frac{d k^i}{d \lam} = 2 \frac{d \Psi}{d \lam} k^i - \left( \de^i \Psi + \de^i \Phi \right) \del_{jk} k^j k^k \, .
\end{equation}
To find the solution of this equation we first decompose $k^i = k_{||} e^i + k^i_{\perp}$ where
$k_{||} = e_j k^j$ and $k^i_\perp = (\del^i_j - e^i e_j) k^j$ are the parallel and transverse parts of the wavevector 
with respect to the background geodesic. Then, in the geodesic equation we approximate 
$\del_{ij} k^j k^k = k_{||}^2 + \del_{jk} k_\perp^j k_\perp^k \simeq k_{||}^2$,
since the second term will be negligible on large angular scales. Then, projecting Eq.(\ref{gsnp}), we can split it in the two equations
\begin{align}
\frac{d k_{||}}{d \lam} =& 2 \frac{d \Psi}{d \lam} k_{||} - e_i \left( \de^i \Psi + \de^i \Phi \right) k_{||}^2 \, , \\
	\frac{d k_\perp^i}{d \lam} =& 2 \frac{d \Psi}{d \lam} k_\perp^i
		- (\del^i_j - e^i e_j) \left( \de^j \Psi + \de^j \Phi \right) k_{||}^2 \, ,
\end{align}
which can be solved. The solution for $k_{||}$ is analogous to the one for $k^0$
\begin{equation}
\label{solkp}
k_{||} = e^{2 \Psi} \left[ - 1 + \int_{\lam_\O}^\lam d \tilde{\lam} e^{2 \Psi}
	e^i \de_i \left( \Psi + \Phi \right) \right]^{-1} \, .
\end{equation}
from which the solution for $k^i_{\perp}$ is easily obtained
\begin{equation}
\label{solkt}
k_\perp^i = - e^{2 \Psi} \int_{\lam_\O}^\lam d \tilde{\lam} e^{2 \Psi}
	(\del^{ij} - e^i e^j) \de_j \left( \Psi + \Phi \right)
	\left[ - 1 + \int_{\lam_\O}^{\tilde{\lam}} d \tilde{\lam} e^i \de_i \left( \Psi + \Phi \right) \right]^{-2} \, ,
\end{equation}
since as initial condition we have $k_\perp^i (\lam_\O) = 0$. Perturbing Eqs.~(\ref{solkp}) and~(\ref{solkt}) up to second-order we find 
\begin{equation}
\label{ki1}
k^i_{(1)} = - 2 \1{\Psi}e^i - \int_{\lam_\O}^\lam d \tilde{\lam} \de^i \1{A}\, ,
\end{equation}
\begin{equation}
\begin{split}
\label{ki2}
k^i_{(2)} =& - \2{\Psi} e^i - 2 x_{(1)}^\mu \de_\mu \1{\Psi} e^i - 2 \1{\Psi}^2 e^i
	- 2 \1{\Psi} \int_{\lam_\O}^\lam d\tilde{\lam} \, \de^i \1{A} \\
	&+ \int_{\lam_\O}^\lam d\tilde{\lam} \left[ - \frac12 \de^i \2{A} - x^\mu_{(1)} \de_\mu \de^i \1{A}
	- 2 \1{\Psi} \de^i \1{A} - 2 \de^i \1{A} \int_{\lam_\O}^{\tilde{\lam}} d\lam' e^j \de_j \1{A} \right]\, .
\end{split}
\end{equation}
Notice that in fact the above expressions for $k^{i}$ are exact, as we have checked explicitly by using the pertubative method of Eqs.~(\ref{fterms})-(\ref{fterms3}).
The reason is that up to this order the large-scale approximation we used also corresponds to neglect higher-order terms in the perturbations. 
Moreover, as a further consistency check, we have explicitly verified that the expressions~\eqref{k01}-\eqref{k03} and~\eqref{ki1}-\eqref{ki2}
are the same one gets with the perturbative method employed in Sec~\ref{2S}.

The integrals~(\ref{I1})-(\ref{I3}) along the line of sight are at the origin of the late (and early) Integrated Sachs-Wolfe effect, since 
they express the redshift the photons suffer from the last scattering surface to the observer while traveling through a time varying gravitational
potential. In particular the first integrand terms, $A^{(2)'}$ and $A^{(3)'}$ in each of the integrals at second and third order, 
$I_2(\lambda)$ and $I_3(\lambda)$, are just the straightforward extension of the well-known first order ISW effect. However one must consider 
all the additional contributions which are integrated terms and which depend on the time variation of the gravitational potentials. 
So terms of this type will appear also from the first two lines of Eq.(\ref{k03}). Notice in particular that terms containing spatial gradients 
like  $x^{(r)j} \partial_j A^{(m)'}$ (and similar) coming from the combination $x^{(r)\mu} \partial_\mu A^{(m)'}=x^{(r)j} \partial_j A^{(m)'}+x^{(r)0} 
\partial_0A^{(m)'}$   cannot be neglected a priori even in the large scale approximation because they are integrated along the line of sight.
\footnote{In fact one can check that for a term like, for example, $x^{(r)j} \partial_jA^{(m)'}$ only the parts of such gradients 
transverse to the background geodesic can be neglected on large scales, while the longitudinal part combines with  $x^{(r)0} \partial_0 A^{(m)'}$ 
to give a term which does not contain spatial gradients.}

Finally plugging the expressions~(\ref{k01})-(\ref{k03}) for $k^{(0)}$ up to third-order into Eq.(\ref{T3scalars})   
the expression for the third order temperature fluctuations on large scales is 
\begin{equation}
\label{T3scalarsls}
\begin{split}
  &\frac{\3{\del} T}{T} = \3{\tau} + \frac{1}{6} \3{\Phi_\E}+
\2{\tau} \1{\Phi_\E} + \1{\tau} \frac{1}{2} \2{\Phi_\E}
 	+ \1{\tau} \frac{1}{2} \1{\Phi_\E}^2+ \frac{1}{2} \1{\Phi_\E} \2{\Phi_\E}+ \frac{1}{6} \left( \1{\Phi_\E} \right)^3\\
& + x^{(1)0}  \2{\Phi_\E}' + \left( x^{(1)0} \right)^2 \1{\Phi_\E}''
 	+ 2  x^{(2)0} \1{\Phi_\E}'  + 2 x^{(1)0} \1{\Phi_\E}' \left(\1{\Phi_\E} + \1{\tau} \right) \\
        &- I_{3\E} - I_{2\E} \left[ \1{\tau} + \1{\Phi_\E} \right]
        - I_{1\E} \left[ \2{\tau} + \1{\tau} \1{\Phi_\E} + \frac{1}{2} \2{\Phi_\E} + \frac{1}{2} \1{\Phi_\E}^2 + 2 x^{(1)0} \1{\Phi_\E}' \right] \\
 	&+ d_{(1)}^i \left[ \frac{\de \2{\tau}}{\de d^i}
 		+ \left( \1{\Phi_\E} - I_{1\E} \right) \frac{\de \1{\tau}}{\de d^i} \right] 
+ \frac{1}{2} d_{(1)}^i d_{(1)}^j \frac{\de^2 \1{\tau}}{\de d^i \, \de d^j} + d_{(2)}^i \frac{\de \1{\tau}}{\de d^i} \\
 	&+ \1{x}^0 \Bigg[ \partial_0\left(-\2{\Phi_\E} + 2 \1{\Phi_\E}^2 - 2 x^{(1)0} \1{\Phi_\E}' \right)
 		+ \frac{1}{2} \2{A_\E}' + x^{(1)0}  \1{A_\E}'' - 4 \1{\Phi_\E} \1{A_\E}' \\
 		&\quad \qquad + I_{1\E} \1{A_\E}'
 		+ I_{1\E} \left( - 2 \1{\Phi_\E}'+ \1{A_\E}' \right)
 		- \frac{1}{2} \1{x}^0 \left( \1{A_\E}''- 2 \1{\Phi_\E}'' \right) \\
 		& \qquad \quad - \left( 2 I_{1\E} - 3 \1{\Phi_\E} - \1{\tau} \right)
 		\left( \1{A_\E}' - 2 \1{\Phi_\E}' \right) \Bigg]
 	+ 2 \2{\Del \lam} \1{\Phi_\E}' - \2{\Del \lam} A_\E^{(1)'}\, ,
\end{split}
\end{equation}
where $A^{(r)}$ is given by Eq.~(\ref{An}) and we recall that $\2{\Del \lam} = - x^{(2)0} + k^{(1)0} x^{(1)0}$. We recognize various contributions to the 
temperature anistropies in Eq.~(\ref{T3scalarsls}).
The first line includes the Sachs-Wolfe effect, which combines the intrinsic temperature fluctuations
$\tau$ with the photon redshift on the last scattering surface due to the gravitational potential perturbations ${\Phi_\E}$. 
The terms in the third line correspond to the Integrated  Sachs-Wolfe effect, while in the fourth line there are contributions due to a lensing effect 
at the last scattering surface, being dependent on the vector ${\bf \hat{d}}$ which specifies the direction of emission of the photon. 
Finally all the remaining terms are due to a possible time dependence of the 
gravitational potentials at the last scattering epoch, due to the fact that by that time the universe is still not completely 
matter dominated (similarly to the early integrated Sachs-Wolfe effect). At second-order, in Eq.~(\ref{T2scalars}), the lensing like 
contributions appear as well. Notice that in order to make a comparison with the bispectrum computed using the different technique of Ref.~\cite{CZRec}, 
one has to take into account also these lensing terms along with the Sachs-Wolfe effect.

\section{Third-order scalar perturbations of a flat $\Lambda$CDM Universe}
\label{e3p}
\subsection{Evolution of the gravitational potentials on large scales}
We now consider a spatially flat Universe filled with a cosmological constant $\Lam$
and a non-relativistic perfect pressureless fluid and we solve the third order Einstein equations for the scalar perturbations,
in the large scale limit. The energy momentum tensor of the matter component reads $T^\mu_\nu = \rho u^\mu u_\nu$, 
with energy density $\rho$ and four-velocity $u^\mu$.
The derivation of the relevant equations is sketched in the Appendix.

The evolution equation for the third order gravitational potential $\3{\psi}$ reads
\begin{equation}
\label{psi}
    \psi''_{(3)} + 3 \H \psi'_{(3)} + a^2 \Lam \psi_{(3)} = S(\eta, \x) \, ,
\end{equation}
where the source term is given by
\begin{equation}
\label{source}
 \begin{split}
  S(\eta, \x) =& \frac{4}{\H^2 \Om_m}
  	\nab^{-2} \Big\{ \left[ \left( \H^2 + a^2 \Lam \right) E + \H E' \right]
  	\left[ 3 \nab^{-2} \de_i \de^j \left( \vphi_0 \de^i \vphi_0 \de_j \vphi_0 \right)
  		- \vphi_0 (\de^l \vphi_0) (\de_l \vphi_0) \right] \\
  	&\qquad + \left[ \left( \H^2 + a^2 \Lam \right) F + \H F' \right]
  		\left[ 3 \nab^{-2} \de_i \de^j \left( \de^i \al_0 \de_j \vphi_0 + \de^i \vphi_0 \de_j \al_0 \right)
  		- 2 (\de^l \al_0) (\de_l \vphi_0) \right] \\
  	&\qquad - \left( g' + 2 \H g \right) \left[ 3 \H (g')^2 + a^2 \Lam g^2 (2 f -1) \right] \vphi_0 \al_0 \Big\} \\
  	&- 6 (g')^2 \vphi_0^3 + 3 \left( g' A' \vphi_0^3 + g' B' \vphi_0 \al_0 \right) + \textrm{gradients} \, ,
 \end{split}
\end{equation}
where $\nabla^{-2}$ stands for the inverse of the Laplacian operator. 
Here $g(\eta)$ is the linear growth function, defined by $\vphi(\x,\eta) = g(\eta) \vphi_0(\x)$,
where $\vphi_0(\x)$ is the peculiar gravitational potential linearly extrapolated to the present time, 
and $f(\eta) = 1 + \frac{g'(\eta)}{\H g(\eta)}$. 
The growth-suppression factor is given by $g(\eta)=D_+(\eta)/a(\eta)$ where $D_+(\eta)$ is the linear growing-mode 
of density fluctuations in the Newtonian limit.
The exact form of $g$ can be found in Refs.~\cite{lahav,Carroll,Eisenstein}.
A very good approximation for $g$ as a function of redshift $z$ is given in Refs.~\cite{lahav,Carroll} 
\begin{equation}
g \propto \Omega_m\left[\Omega_m^{4/7} - \Omega_\Lambda +
\left(1+ \Omega_m/2\right)\left(1+ \Omega_\Lambda/70\right)\right]^{-1} \;, 
\end{equation}
with $\Omega_m=\Omega_{0m}(1+z)^3/E^2(z)$, 
$\Omega_\Lambda=\Omega_{0\Lambda}/E^2(z)$, 
$E(z) \equiv (1+z) {\mathcal H}(z)/{\mathcal H}_0 = \left[\Omega_{0m}(1+z)^3 + 
\Omega_{0\Lambda}\right]^{1/2}$ and 
$\Omega_{0m}$, $\Omega_{0\Lambda}=1-\Omega_{0m}$, the present-day
density parameters of non-relativistic matter and cosmological constant, 
respectively. We will normalize the growth-suppression factor so that 
$g(z=0)=1$. The function $f(\eta)$ can be written as a function of $\Omega_m$ as $f(\Omega_m) 
\approx \Omega_m(z)^{4/7}$ ~\cite{lahav,Carroll}. In the $\Lambda=0$ case $g=1$ and $f(\eta)=1$.

In Eq.(\ref{source}) $A(\eta)$, $B(\eta)$ enter in the solutions for the second order gravitational potentials 
on large scales \cite{fullT}:
\begin{align}
 \2{\Psi}(\x,\eta) =& A(\eta) \vphi_0^2(\x) + B(\eta) \al_0(\x) \, \\
 \2{\Phi}(\x,\eta) =& A(\eta) \vphi_0^2(\x) + C(\eta) \al_0(\x) \,
\end{align}
and their explicit expressions are
\begin{align}
  \label{al0} \al_0(\x) =& \nab^{-2} \left[ \de^i \vphi_0 \de_i \vphi_0
      - 3 \nab^{-2} \de_i \de^j (\de^i \vphi_0 \, \de_j \vphi_0) \right] \, , \\
  \label{A} A(\eta) =& B_1(\eta) + 2 g^2(\eta) - 2 g_m g(\eta)
      - \frac{10}{3} \left( a_{NL} - 1 \right) g_m g(\eta) \, , \\
  \label{B} B(\eta) =& B_2(\eta) - \frac{4}{3} g_m g(\eta) \, , \\
  \label{C} C(\eta) =&  B(\eta) + \frac{4}{3} g_m g(\eta) \left( \frac{f^2(\eta)}{\Om_m(\eta)} + \frac{3}{2} \right) \, ,
\end{align}
where the functions $B_1(\eta)$ and $B_2(\eta)$ are given in Ref.~\cite{fullT} while $g(\eta_m)=g_m$ is the value of the growth suppression 
factor during matter domination, when the cosmological constant was still negligible. A good approximation is 
$g_m \approx \frac{2}{5} \Omega_{0m}^{-1}(\Omega_{0m}^{4/7} + \frac{3}{2}\Omega_{0m})$.  

Notice that in $A(\eta)$ appears the parameter $a_{\rm NL}$ specifying the 
level of quadratic primordial non-Gaussianity which depends on the particular scenario for the generation of cosmological perturbations, as we will 
discuss 
later in more detail. For example, for standard single-field models of slow-roll inflation $a_{\rm NL}=1+{\cal O}(\epsilon, \eta)$
~\cite{ABMR,maldacena,BMR2,review}, where $\epsilon$ and $\eta$ are the standard slow-roll parameters~\cite{lrreview}. The reason it 
appears in Eq.~(\ref{source}) is due to the fact that at third-order some of the source terms contain products of linear and  
second-order gravitational potentials, as detailed in the Appendix.

We have also introduced the functions
\begin{align}
  \begin{split}
  \label{E}
  E(\eta) =& 2 A' g' + \left( 5 \H^2 - a^2 \Lam \right) A g - 2 \H g' g^2
  + 2 \H (A' g + g' A) - 2 g (g')^2 \\
  &- 2 g (g' + \H g)^2 + 2 \frac{g f}{\Om_m} \left( g' + \H g \right)^2 \, ,
  \end{split}\\
  \label{F}
  F(\eta) =& \frac{1}{2} B' g' + \frac{1}{4} \left( 5 \H^2 - a^2 \Lam \right) C g
        + \frac{1}{2} \H (C g' + B' g) \, .
\end{align}
Notice that when $\Lambda=0$ then $B_1(\eta)$ and $B_2(\eta)\rightarrow 0$,$B(\eta)\rightarrow-4/3$ and  
$A(\eta)\rightarrow 2 -(10/3) a_{\rm NL}$, and therefore $E(\eta)$ simplifies to $E(\eta)=-(50/3) a_{\rm NL} {\cal H}^2$ while 
$F(\eta)=(5/2){\cal H}^2$. Thus, looking at Eq.(\ref{source}), it is simple to see that all the terms in the last two lines are specific only to 
the case of a non-vanishing cosmological constant.  

The solution of Eq.\eqref{psi} is given by the Green's formula
\begin{equation}
\label{green}
 \3{\psi}(\eta) = \frac{g}{g_m} \3{\psi_m}
 	+ \psi_+ (\eta) \int_{\eta_m}^\eta d \eta' \frac{\psi_- (\eta')}{W(\eta')} S(\eta')
 	+ \psi_- (\eta) \int_{\eta_m}^\eta d \eta' \frac{\psi_+ (\eta')}{W(\eta')} S(\eta') \, ,
\end{equation}
where
\begin{equation}
 \psi_+ (\eta) = g(\eta) \, , \qquad
 \psi_- (\eta) = \frac{\H(\eta)}{a^2(\eta)} \, , \qquad
 W(\eta) = \frac{\H_0^2}{a^3(\eta)} \left( f_0 + \frac{3}{2} \Om_{0m} \right) \, ,
\end{equation}
are respectively the growing and decaying mode solutions and the Wronskian of the homogeneous equation. 
Here the suffix `0' stands for the value of the corresponding quantities at the 
present time, while $\psi^{(3)}_{m} \equiv \psi^{(3)}(\eta_m)$ represents the 
initial condition taken deep in the matter dominated era on super-horizon 
scales, $\eta_m$ being the epoch when full matter domination starts. It is such an initial value 
that must be properly determined in order to account for the primordial cubic non-Gaussianity in the cosmological perturbations.

The solution for $\3{\phi}$ is then obtained from the relation between $\3{\psi}$ and $\3{\phi}$ obtained in the Appendix (Eq.~\ref{tracelessnp})
\begin{equation}
\label{relationphipsi}
 \3{\phi} = \3{\psi} + 12 g(\eta) A(\eta) \vphi_0^3 + 6 g(\eta) \left( B(\eta) + C(\eta)\right) \al_0 \vphi_0
 	- \frac{4}{\H^2 \Om_m} \left[ E(\eta) \mu_0 + F(\eta) \nu_0 \right] \, ,
\end{equation}
where we have defined the following kernels:
\begin{align}
 \mu_0 (\x) =& 3 \nab^{-4} \de_i \de^j \left(\vphi_0 \, \de^i \vphi_0 \, \de_j \vphi_0 \right)
        - \nab^{-2} \left( \vphi_0 \, \de^k \vphi_0 \, \de_k \vphi_0 \right) \, ,\\
 \nu_0 (\x) =& 3 \nab^{-4} \de_i \de^j \left( \de^i \al_0 \, \de_j \vphi_0 + \de^i \vphi_0 \de_j \al_0 \right)
       - 2 \nab^{-2} (\de^k \al_0 \, \de_k \vphi_0) \, .
\end{align}

\subsection{Initial conditions from primordial non-Gaussianity}
\label{IC}
We now discuss the key issue of the initial conditions,
which are conveniently fixed at the time when the relevant modes of the perturbations are well
outside the Hubble radius.
In order to follow the super-horizon evolution of the density perturbations produced during inflation,
we use the curvature perturbations on uniform-density hypersurfaces $\zt$,
which will be expanded as $\zt = \1{\zt} + \frac{1}{2} \2{\zt} + \frac{1}{6} \3{\zt} + \dots$, where at linear order 
$\zeta^{(1)}=-\psi^{(1)}-{\cal H} \delta^{(1)} \rho/\bar{\rho}$ was first introduced in Ref.~\cite{BST}. For our purposes it is useful  
to adopt a non-perturbative generalization of $\zt$ given by (see Refs.~\cite{SalopekBond,KMNR,Bartolo:2005fp})
\begin{equation}
\label{znp}
 \zt = - \Psi + \frac{1}{3(1+w)} \ln \frac{\rho}{\bar{\rho}} \, ,
\end{equation}
where just for simplicity in this expression we have assumed a constant equation of state parameter $w=\bar{p}/\bar{\rho}$, 
$\bar\rho$ and $\bar{p}$ being the background energy density and pressure respectively. 
This variable has two crucial features: it is \emph{gauge invariant} and, most important, it is \emph{constant} in time
for adiabatic, super-horizon perturbations.\footnote{
As a consistency test, we have used Eq.~(\ref{znp}) together with the Einstein equations contained in the Appendix, 
to verify that $\zeta'$ turns out to be proportional to the L.H.S. of Eq.~(\ref{psi}) minus the very same source term 
appearing in Eq.~(\ref{source}). This provides an independent check for the expression of the source term which turns out 
to be consistent with $\zeta'=0$ on large scales.
}
This allows us to set the initial conditions at the time when $\zt$ becomes constant,
and follow them until the perturbation mode reenters the horizon.
The primordial non-Gaussianity can be parametrized in terms of the curvature perturbation as~\cite{Bartolo:2005fp}
\begin{equation}
 \2{\zt} = 2 (a_{\rm NL}-1) \left( \1{\zt} \right)^2 \, , \qquad \3{\zt} = 6 (b_{\rm NL}-1) \left( \1{\zt} \right)^3 \, ,
\end{equation}
where the two non-linearity parameters depend on the physics of a given scenario of generation of the perturbations. For example for standard 
single-field models of inflation $a_{\rm NL}=1$ and $b_{\rm NL}=1$ (plus tiny corrections proportional to the slow-roll parameters), while for other 
scenarios they might well be non-negligible.   

On the other hand, the physical observable quantity is given by the CMB anisotropies, which are one of the best tool to detect or constrain the 
primordial non-Gaussianity generated on large scales~\cite{review}. 
The standard procedure is to introduce the quadratic and cubic non-linearity parameters, $f_{\rm NL}$ and $g_{\rm NL}$ 
(which can be in fact also non-trivial kernels in Fourier and harmonic space) characterizing the non-Gaussianity in the large scale temperature 
anisotropies~\cite{ks,OkamotoHu,kwmap,LLMR,kwmap3}. In the limit of large non-Gaussianity, $|a_{\rm NL}| \gg 1$ and $|b_{\rm NL}| \gg 1$
~\cite{review,prl,Bartolo:2005fp}, 
the resulting size of the non-Gaussianity in the CMB anisotropies can be estimated as $f_{\rm NL}\sim 5 a_{\rm NL}/3$ and $g_{\rm NL} \sim 
25 b_{\rm NL}/9$ if accounting only for the contribution to the CMB anisotropies from the Sachs-Wolfe effect. In fact for the cubic 
non-linearities entering in the Sachs-Wolfe effect such an estimate is too rough, 
and the correct expression relating the observable quantity $g_{\rm NL}$ to the primordial 
non-linearity parameters $a_{\rm NL}$ and $b_{\rm NL}$ is given by Eq.~(67) of Ref.~\cite{Bartolo:2005fp}. 

Our results allow to take into account also 
the contribution from the (late) Integrated Sachs-Wolfe effect. The memory of the initial non-Gaussianity is kept in the matter-dominated value of the 
large-scale gravitational potential $\psi^{(3)}_m$, and also, as noticed before, in the source term~(\ref{source}). In order to determine  
$\psi^{(3)}_m$ we use the expression~(\ref{znp}) for the curvature perturbation in the matter dominated era 
\begin{equation}
 \zt_m = - \Psi_m - \frac{2}{3} \Phi_m \, ,
\end{equation}
where we have employed the energy constraint~(\ref{00}) to replace for the energy density. 

We can expand both sides: at first order we find the usual result
\begin{equation}
 \1{\zt_m} = - \frac{5}{3} g_m \vphi_0 \, ;
\end{equation}
at second order
\begin{equation}
 \2{\phi_m} = - \frac{3}{5} \2{\zt_m} + 2 g_m^2 \left( \vphi_0^2 + \al_0 \right)
	= 2 g_m^2 \left[  - \frac{5}{3} (a_{\rm NL}-1) \vphi_0^2 + \vphi_0^2 +\al_0 \right] \, ;
\end{equation}
and at third order
\begin{equation}
 \3{\phi_m} = - \frac{3}{5} \3{\zt_m} + 6 \vphi_m \2{\phi_m} - 8 \vphi_m^3
 	+ \frac{12}{5 \H^2 \Om_m} \left[ E(\eta) \mu_0 + F(\eta) \nu_0 \right] \, .
\end{equation}

With the use of Eq.~\eqref{relationphipsi} evaluated in the matter dominated period, we get as initial conditions
\begin{align}
\label{Phiba}
 \3{\phi_m} =& 6 g_m^3 \left[ \left(\frac{25}{9} (b_{\rm NL}-1) - \frac{10}{3} (a_{\rm NL}-1) + \frac{2}{3} \right) \vphi_0^3
 	+ 2 \vphi_0 \al_0 - \frac{20}{3} (a_{\rm NL}-1) \mu_0 + \nu_0 \right] \, , \\
 \3{\psi_m} =& 6 g_m^3 \left[ \left(\frac{25}{9}(b_{\rm NL}-1)  + \frac{10}{3} (a_{\rm NL}-1)  + \frac{2}{3} \right) \vphi_0^3
 	- \frac{4}{3} \vphi_0 \al_0 - \frac{160}{9}  (a_{\rm NL}-1) \mu_0 + 8 \nu_0 \right] \, .
\end{align}

Finally, from Eqs.~\eqref{green}, \eqref{relationphipsi}, the large-scale solutions for $\3{\psi}$, $\3{\phi}$ are given by 
\begin{align}
\begin{split}
 &\3{\psi}(\eta,\x) = 6 g(\eta) g_m^2 \left[ \left(\frac{25}{9} (b_{\rm NL}-1)  + \frac{10}{3} (a_{\rm NL}-1)
		+ \frac{2}{3} \right) \vphi_0^3(\x)
 	- \frac{4}{3} \vphi_0(\x) \al_0(\x) - \frac{160}{9} (a_{\rm NL}-1) \mu_0(\x) + 8 \nu_0(\x) \right] \\
	&+ \H_0^{-2} \left( f_0 + \frac{3}{2} \Om_{0m} \right)^{-1}
	\left[ g(\eta) \int_{\eta_m}^\eta d\t{\eta} a(\t{\eta}) \H(\t{\eta}) S(\t{\eta},\x)
	- \frac{\H(\eta)}{a^2(\eta)} \int_{\eta_m}^\eta d\t{\eta} a^3(\t{\eta}) g(\t{\eta}) S(\t{\eta},\x)\right] \, ;
\end{split}\\
\begin{split}
 &\3{\phi}(\eta,\x) = \3{\psi}(\eta,\x) + 12 g(\eta) A(\eta) \vphi_0^3(\x)
 	+ 6 g(\eta) \left(B(\eta) + C(\eta) \right) \al_0(\x) \vphi_0(\x) \\
 	&- \frac{4}{\H^2 \Om_m} \left[E(\eta) \mu_0(\x) + F(\eta) \nu_0(\x) \right] \, .
\end{split}
\end{align}

\section{Conclusions}
In this paper we have computed the expression for the CMB anisotropies due to the 
redshift the photons suffer when they travel from the last scattering surface to the observer up to third-order in the gravitational 
perturbations. 
We thus have completed the findings of Ref.~\cite{Bartolo:2005fp}, where a fully non-linear expresssion for the Sachs-Wolfe effect 
has been obtained, 
by including also the Integrated Sachs-Wolfe effect and lensing effects 
at the last scattering surface. To achieve  this goal we have proposed an 
alternative method to the standard perturbative one in order to solve for the geodesic photon equation which provides a fully 
non-linear integral solution.
Moreover we have studied the evolution of the gravitational potentials on large scales, allowing for generic non-Gaussian 
initial conditions. Our results, together with those 
of Ref.~\cite{Bartolo:2005fp}, are of particularly relevance when facing the trispectrum of the CMB anisotropies, in that, 
as pointed out in the 
Introduction, they include all the relevant cubic non-linearities for a coherent prediction of such statistic from various 
cosmological scenarios.
In particular they include those contributions related to primordial non-Gaussianity, some of which would be inevitably lost 
if one would stick to the linear evolution 
approximation that is often used in the literature.          

\section*{Acknowledgments}
The authors wish to thank the anonymous Referee for the suggestions which helped to improve the paper.

\newpage

\appendix
\setcounter{equation}{0}

\section{Connection coefficients}

We start from the line element of the conformal metric
\begin{equation}
 ds^2 = - e^{2 \Phi} d\eta^2 + 2 \om_i d\eta dx^i + \left( e^{-2 \Psi} \del_{ij} + \chi_{ij} \right) dx^i dx^j \, .
\end{equation}

It is useful to consider as ``background'' metric the one with the exponentiated scalars:
\begin{equation}
 \bar{g}_{00} = - e^{2 \Phi} \, , \qquad \bar{g}_{0i} = 0 \, , \qquad \bar{g}_{ij} = e^{- 2 \Psi} \del_{ij} \, ,
\end{equation}
while treating vectors and tensors perturbatively (so here $n$ refers to the order in vectors and tensors):
\begin{equation}
 \del g^{(n)}_{00} = 0 \, , \qquad \del g^{(n)}_{0i} = \frac{1}{n!} \om^{(n)}_i \, , \qquad
 	\del g^{(n)}_{ij} = \frac{1}{n!} \chi^{(n)}_{ij} \, .
\end{equation}

The inverse metric will be computed perturbatively up to third order.
At zeroth order
\begin{equation}
 \bar{g}^{00} = - e^{- 2 \Phi} \, , \qquad \bar{g}^{0i} = 0 \, , \qquad \bar{g}^{ij} = e^{2 \Psi} \del^{ij} \, .
\end{equation}

At first order, $\del g_{(1)}^{\mu \nu} = - \bar{g}^{\mu \lam} \bar{g}^{\nu \rho} \del g^{(1)}_{\lam \rho}$ and we find
\begin{equation}
 \del g^{00}_{(1)} = 0 \, , \qquad \del g^{0i}_{(1)} = e^{2 (\Psi -\Phi)} \om_{(1)}^i \, , \qquad
 	\del g^{ij}_{(1)} = - e^{4 \Psi} \chi_{(1)}^{ij} \, .
\end{equation}

At second order, $\del g_{(2)}^{\mu \nu} = - \bar{g}^{\mu \lam} \bar{g}^{\nu \rho} \del g^{(2)}_{\lam \rho}
- \del g_{(1)}^{\mu \lam} \bar{g}^{\nu \rho} \del g^{(1)}_{\lam \rho}$ and we find
\begin{align}
 \del g^{00}_{(2)} =& e^{2 \Psi - 4 \Phi} \left( \1{\om_k} \right)^2 \nonumber \\
 \del g^{0i}_{(2)} =& e^{2 (\Psi - \Phi)} \left[ \frac{1}{2} \om_{(2)}^i - e^{2\Psi} \1{\om_k} \chi_{(1)}^{ki} \right] \\
 \del g^{ij}_{(2)} =& e^{4 \Psi} \left[ - \frac{1}{2} \chi_{(2)}^{ij} - e^{-2\Phi} \om_{(1)}^i \om_{(1)}^j
    + e^{2\Psi} \chi_{(1)}^{il} \chi^{(1)j}_l \right] \nonumber \, .
\end{align}

At third order, $\del g_{(3)}^{\mu \nu} = - \bar{g}^{\mu \lam} \bar{g}^{\nu \rho} \del g^{(3)}_{\lam \rho}
- \del g_{(1)}^{\mu \lam} \bar{g}^{\nu \rho} \del g^{(2)}_{\lam \rho}
- \del g_{(2)}^{\mu \lam} \bar{g}^{\nu \rho} \del g^{(1)}_{\lam \rho}$:
\begin{align}
 \del g^{00}_{(3)} =& e^{2 \Psi - 4 \Phi} \left[ \om_{(1)}^i \2{\om_i}
 	- e^{2 \Psi} \1{\om_i} \1{\om_j} \chi_{(1)}^{ij} \right] \, , \nonumber \\
 \del g^{0i}_{(3)} =& e^{2(\Psi-\Phi)} \left[ \frac{1}{6} \om_{(3)}^i - \frac{1}{2} e^{2\Psi} \1{\om_k} \chi_{(2)}^{ki}
   - e^{2(\Psi-\Phi)} \left(\1{\om_l}\right)^2 \om_{(1)}^i
        - \chi_{(1)}^{ik} e^{2 \Psi} \left( \frac{1}{2} \om_{(2)}^k - e^{2\Psi} \1{\om_l} \chi_{(1)}^{lk} \right) \right] \, , \nonumber \\
 \del g^{ij}_{(3)} =& e^{4\Psi} \left[ - \frac{1}{6} \chi_{(3)}^{ij}
        - \frac{1}{2} e^{-2\Phi} \left( \om_{(1)}^i \om_{(2)}^j + \om_{(2)}^i \om_{(1)}^j \right)
        + \frac{1}{2} e^{2\Psi} \left( \chi_{(1)}^{il} \chi^{(2)j}_l + \chi_{(2)}^{il} \chi^{(1)j}_l \right) \right. \\
        &\left. \qquad + e^{2(\Psi-\Phi)} \1{\om_l} \left( \chi_{(1)}^{li} \om_{(1)}^j
        	+ \chi_{(1)}^{lj} \om_{(1)}^i \right)
        - e^{4\Psi} \chi_{(1)}^{ik} \1{\chi_{kl}} \chi_{(1)}^{lj} \right] \nonumber \, .
\end{align}

It is now convenient to resum the inverse metric,
retaining terms up to third order; we find
\begin{align}
\label{resummed}
 g^{00} =& - e^{-2\Phi} + e^{2\Psi-4\Phi} \left(\om_k \right)^2 - \om^i \om^j \chi_{ij} \nonumber \, , \\
 g^{0i} =& e^{2(\Psi-\Phi)} \om^i - e^{4\Psi-2\Phi} \om_k \chi^{ki} - (\om_k)^2 \om^i + \chi^{ij} \chi_{jk} \om^k \, ,\\
 g^{ij} =& e^{2\Psi} \del^{ij} - e^{4\Psi} \chi^{ij} - e^{4\Psi-2\Phi} \om^i \om^j + e^{6\Psi} \chi^{ik} \chi_k^j
    + \om_k \left( \om^i \chi^{kj} + \om^j \chi^{ki} \right) - \chi^{ik} \chi_{kl} \chi^{lj} \nonumber \, .
\end{align}

It is now immediate to compute the connections coefficients with the the usual formula
\begin{equation}
 \Gam^\mu_{\al \bt} =
 \frac12 g^{\mu \lam} \left( \de_\al g_{\bt \lam} + \de_\bt g_{\al \lam} - \de_\lam g_{\al \bt} \right) \, ,
\end{equation}
where $g^{\mu \nu}$ is the resummed inverse metric in eq.\eqref{resummed}.
We find, keeping terms up to third order:
\begin{equation}
 \Gam^0_{00} = \Phi' \left[ 1 - (\om_k)^2 \right]
        + \left( e^{2\Phi} \de_i \Phi + \om'_i \right) \left[ e^{2(\Psi-\Phi)} \om^i - \chi^{ij} \om_j \right] \, ;
\end{equation}

\begin{equation}
 \Gam^0_{0i} = \de_i \Phi \left[ 1 - (\om_k)^2 \right]
    + \Psi' \left[ - e^{-2 \Phi} \om_i + \chi_{ij} \om^j \right]
    + \frac12 \left[ \chi'_{ij} + \de_i \om_j - \de_j \om^i \right]
        \left[ e^{2(\Psi-\Phi)} \om^j - \chi^{jk} \om_k \right] \, ;
\end{equation}

\begin{equation}
 \begin{split}
  \Gam^0_{ij} =& \Psi' \del_{ij} \left[ - e^{-2(\Psi+\Phi)} + (\om_k)^2 \right] + 
            \frac12 \left[ \chi'_{ij} - (\de_i \om_j + \de_j \om_i) \right] \left[ e^{-2\Phi} - (\om_k)^2 \right] \\
        & + \frac12 \left(\de_i \chi_{jk} + \de_j \chi_{ik} - \de_k \chi_{ij} \right)
        \left[ e^{2(\Psi-\Phi)} \om^k - \om_l \chi^{lk} \right] \\
        & + \left( \de_i \Psi \del_{jk} + \de_j \Psi \del_{ik} - \de_k \Psi \del_{ij} \right)
        \left[ - e^{-2\Phi} \om^k + \om_l \chi^{lk} \right] \, ;
 \end{split}
\end{equation}

\begin{equation}
  \Gam^i_{00} = \Phi' \left[ - e^{2\Psi} \om^i + \om_j \chi^{ij} \right]
        + \left( \om'_j + e^{2\Phi} \de_j \Phi \right) \left[ e^{2\Psi} \del^{ij} - e^{4\Psi} \chi^{ij}
            - \om^i \om^j + \chi^{ik} \chi_k^j \right] \, ;
\end{equation}

\begin{equation}
\begin{split}
  \Gam^i_{0j} =& \de_j \Phi \left[ - e^{2\Psi} \om^i + \om_k \chi^{ik} \right] \\
        &+ \frac12 \left( \chi'_{jk} - 2 \Psi' e^{-2\Psi} \del_{jk} + \de_j \om_k - \de_k \om_j \right)
         \left[ e^{2\Psi} \del^{ik} - e^{4\Psi} \chi^{ik} - \om^i \om^k + \chi^{il} \chi_l^k \right] \, ;
\end{split}
\end{equation}

\begin{equation}
 \begin{split}
  \Gam^i_{jk} =& \frac12 \left[ \left( \de_j \om_k + \de_k \om_j \right)
            + 2 \Psi' e^{-2\Psi} \del_{jk} - \chi'_{jk} \right]
            \left[ e^{2(\Psi-\Phi)} \om^i - \om_k \chi^{ik} \right] \\
        &+ \frac12 \left[ \left(\de_j \chi_{kl} + \de_k \chi_{jl} - \de_l \chi_{jk} \right)
            - 2 e^{-2 \Psi} \left(\de_j \Psi \del_{kl} + \de_k \Psi \del_{jl} - \de_l \Psi \del_{jk} \right) \right]
            \times \\
        &\quad \times \left[ e^{2\Psi} \del^{il} - e^{4\Psi} \chi^{il}
            - e^{4\Psi-2\Phi} \om^i \om^l + e^{6\Psi} \chi^{ir} \chi_r^l \right] \, .
 \end{split}
\end{equation}

\section{Einstein equations for the gravitational potentials}

\setcounter{equation}{0}

In this Appendix we derive the evolution equations for the third order scalar perturbations $\3{\psi}$, $\3{\phi}$.

We start by writing the fully non-linear Einstein equations for the metric
$ds^2 = a^2(\eta) \left[- e^{2 \Phi} d\eta^2 + e^{-2 \Psi} \del_{ij} dx^i dx^j \right]$,
for a $\Lam$CDM model:
\begin{itemize}
 \item{0-0} 
 \begin{equation}
\label{00}
 e^{-2 \Phi} \left( \H - \Psi' \right)^2 = \frac{a^2}{3} \left( 8 \pi G \rho + \Lam \right) \, ;
 \end{equation}
 \item{0-i}
 \begin{equation}
  e^{-\Phi} \left[\de_i \Psi'  + \left(\H - \Psi' \right) \de_i \Phi \right] = - 4 \pi G a \rho u_i \, ;
 \end{equation}
 \item{i-j traceless}
 \begin{equation}
 \begin{split}
  &\de^i \de_j \Psi - \frac13 \nab^2 \Psi \del^i_j - \de^i \de_j \Phi  + \frac13 \nab^2 \Phi
  + \de^i \Psi \de_j \Psi - \frac13 \de^l \Psi \de_l \Psi \del^i_j  \\
  & - \de^i \Phi \de_j \Phi + \frac13 \de^l \Phi \de_l \Phi \del^i_j - \de^i \Phi \de_j \Psi
  - \de^i \Psi \de_j \Phi + \frac23 \de^l \Phi \de_l \Psi \del^i_j = \\
  =& 8 \pi G a^2 e^{- 2 \Psi} \left( T^i_j - \frac{1}{3} \del^i_j T \right) \, ;
 \end{split}
 \end{equation}
 \item{i-j trace}
 \begin{equation}
 \begin{split}
  & e^{- 2 \Phi} \left[ \left( \H-\Psi' \right)
    \left( -\H + 3\Psi' + 2\Phi' \right) - 2 \H' + 2 \Psi'' \right] \\
  &+ \frac{e^{2 \Psi}}{3} \left[ 2 \nab^2 \left(\Phi - \Psi \right) + \de^l \Psi \de_l \Psi
   + 2 \de^l \Phi \de_l \Phi - 2 \de^l \Phi \de_l \Psi \right] =\\
  =& \frac{8 \pi G}{3} a^2 T \, .
\end{split}
 \end{equation}
\end{itemize}
In writing down the first two equations, we made use of the normalization condition for the velocity,
$g^{\mu \nu} u_\mu u_\nu = -1$, which gives
\begin{equation}
u_0 = - e^\Phi \left( a^2 + e^{2\Psi} u^l u_l \right)^{\frac12} \simeq - a e^{\Phi} \, ;
\end{equation}
the last equality is valid at any order on large scales,
and up to third order on smaller scales.

We start from the traceless equation, which we project onto the scalar modes by applying the operator $\de_i \de^j$:
\begin{equation}
\begin{split}
  \nab^2 \nab^2 &(\Psi - \Phi)
  = - \frac12 \nab^2 \left[ (\de^l \Psi) (\de_l \Phi) - (\de^l \Psi) (\de_l \Psi)
    + 2 (\de^l \Phi) (\de_l \Psi) + 8 \pi G a^2 e^{-2\Psi} T \right] \\
  &+ \frac{3}{2} \de_i \de^j \left[ (\de^i \Phi) (\de_j \Phi) - (\de^i \Psi) (\de_j \Psi)
    + (\de^i \Phi) (\de_j \Psi) + (\de^i \Psi) (\de_j \Phi) + 8 \pi G a^2 e^{-2\Psi} T^i_j \right] \, .
\end{split}
\end{equation}

It is convenient to rewrite it as
\begin{equation}
\label{tracelessnp}
  \Psi - \Phi = \Q \, ,
\end{equation}
where we define
\begin{equation}
\label{Qnp}
  \nab^2 \Q \equiv - P + 3 N \equiv - P^l_l + 3 \nab^{-2} \de_i \de^j P^i_j
\end{equation}
and
\begin{equation}
  P^i_j \equiv \frac12 \left[ (\de^i \Phi) (\de_j \Phi) - (\de^i \Psi) (\de_j \Psi)\right]
    + \frac12 \left[ (\de^i \Phi) (\de_j \Psi) + (\de^i \Psi) (\de_j \Phi) \right]
    + 4 \pi G a^2 e^{-2\Psi} T^i_j \, .
\end{equation}

For a $\Lam$CDM model, the spatial part of the stress-energy tensor is
\begin{equation}
\begin{split}
  T^i_j =& \frac{e^{2\Psi}}{a^2} \rho u^i u_j - \frac{\Lam}{8\pi G} \del^i_j = \\
  =& \frac{1}{16 \pi^2 G^2} \frac{e^{2(\Psi-\Phi)}}{a^4 \rho}
    \left[ \de^i \Psi' + \left( \H - \Psi'\right) \de^i \Phi \right]
    \left[ \de_j \Psi' + \left( \H - \Psi'\right) \de_j \Phi \right]
    - \frac{\Lam}{8 \pi G} \del^i_j \, ,
\end{split}
\end{equation}
where in the last equality we substitute for the spatial velocities
using the $0-i$ equation:
\begin{equation}
  u_i = - \frac{1}{4 \pi G a} \frac{e^{-\Phi}}{\rho}
    \left[ \de_i \Psi' + \left( \H - \Psi'\right) \de_i \Phi \right] \, .
\end{equation}
 
Substituting this expression into $P^i_j$ we find
\begin{equation}
\begin{split}
  P^i_j =& \frac12 \left[ \de^i \Phi \de_j \Phi - \de^i \Psi \de_j \Psi \right]
    + \frac12 \left[ \de^i \Phi \de_j \Psi + \de^i \Psi \de_j \Phi \right] \\
  &+ \frac{1}{4 \pi G a^2} \frac{e^{-2\Phi}}{\rho}
    \left[ \de^i \Psi' + \left( \H - \Psi'\right) \de^i \Phi \right]
    \left[ \de_j \Psi' + \left( \H - \Psi'\right) \de_j \Phi \right]
    - \frac12 e^{-2\Psi} a^2 \Lam \del^i_j \, .
\end{split}
\end{equation}

Take now the trace equation, and using the background equation $\H^2 + 2 \H' = a^2 \Lam$ and
eqs. \eqref{tracelessnp}, \eqref{Qnp}, we find
\begin{equation}
\label{tracenp}
\begin{split}
  &e^{-2 \Phi} \left[ -a^2 \Lam - 2 \Psi' \Phi' - 3 (\Psi')^2 + 2\H (3 \Psi' - \Q')  + 2 \Psi''\right] \\
  &+ \frac{e^{2\Psi}}{3} \left[ 2 \de^l \Phi \de_l \Phi + \de^l \Psi \de_l \Psi +
    - 2 \de^l \Phi \de_l \Psi + 2 (P - 3 N) \right]
  = \frac{8 \pi G}{3} a^2 T \, .
\end{split}
\end{equation}

At this point, we can expand eq. \eqref{tracenp} at third order.
Using the first order solution $\1{\Phi} = \1{\Psi} = \vphi$, we find:
\begin{equation}
  \begin{split}
    & \3{\Psi}'' + 3 \H \3{\Psi}' + a^2 \Lam \3{\Psi} =
    6 a^2 \Lam \3{\Q} + 6 a^2 \Lam \vphi \2{\Phi}
    - 4 a^2 \Lam \vphi^3 - 30 \vphi (\vphi')^2 + 3 \vphi' (4 \2{\Psi}' + \2{\Phi}') \\
    &+ 18 \H \2{\Phi} \vphi' - 36 \H \vphi^2 \vphi' + 18 \H \vphi \2{\Psi}' 
    - 12 \H \vphi \2{\Q}' + 6 \H \3{\Q}' \\
    &+ 6 \2{\Phi} \vphi'' - 12 \vphi^2 \vphi'' + 6 \vphi \2{\Psi}'' 
    - 2 \vphi (\de_k \vphi)^2 - \de^l \2{\Phi} \de_l \vphi \\
    &+ 4 \vphi \nab^2 \2{\Q} + 2 \nab^2 \3{\Q} + 8 \pi G a^2 \3{T} \, ,
  \end{split}
\end{equation}
where $\Q = \1{\Q} + \2{\Q} + \3{\Q} + \dots$.

Since we are interested in the large scale solution, it is more convenient
to use the small letter variables $\psi^{(r)}$, $\phi^{(r)}$,
related to the capital letter ones as explained after eq. \eqref{metricdef}.
Finally, using the first order equation $\vphi'' + 3 \H \vphi' + a^2 \Lam \vphi = 0$, we get
\begin{equation}
  \begin{split}
    & \3{\psi}'' + 3 \H \3{\psi}' + a^2 \Lam \3{\psi} = \\
    =& 6 a^2 \Lam \3{\Q} + 6 \H \3{\Q}'
    -  6 \vphi (\vphi')^2 + 3 \vphi' \2{\Psi}' - 6 \vphi' \2{\Q}'
    - 12 \H \vphi \2{\Q}' - 2 \vphi (\de_k \vphi)^2 - (\de^l \2{\Phi} \de_l \vphi) \\
    &+ 4 \vphi \nab^2 \2{\Q} + 2 \nab^2 \3{\Q} + 8 \pi G a^2 \3{T} \, .
  \end{split}
\end{equation}

At this point we need to explicit the terms $\3{\Q}$, $\3{\Q}'$ and $\3{T}$.
It is straightforward to show that
\begin{equation}
    \3{T} = \frac{1}{(4\pi G a^2)^2 \rho_0} \left( \de_i \vphi' + \H \de_i \vphi \right)
    \left[ \de_i \2{\Psi}' + \H \de_i \2{\Phi} - 2 \vphi' \de_i \vphi
      - \1{\del} \left( \de_i \vphi' + \H \de_i \vphi \right) \right]
\end{equation}
and
\begin{equation}
  \begin{split}
    P^i_j =& \frac{1}{4 \pi G a^2 \rho_0}
      \Big\{ \frac{1}{2} \left[ \de^i \2{\Psi}' \, \de_j \vphi' + \de^i \vphi' \, \de_j \2{\Psi}' \right]
      + \frac{1}{4} \left( 5 \H^2 - a^2 \Lam \right)
      \left[ \de^i \2{\Phi} \, \de_j \vphi + \de^i \vphi \, \de_j \2{\Phi} \right] \\
      & - 2 \H \vphi' \, \de^i \vphi \, \de_j \vphi
      + \frac{1}{2} \H \left[ \de^i \2{\Phi} \, \de_j \vphi' + \de^i \vphi' \, \de_j \2{\Phi}
        + \de^i \2{\Psi}' \, \de_j \vphi + \de^i \vphi \, \de_j \2{\Psi}' \right] \\
      &- \vphi' \, \left[ \de^i \vphi' \, \de_j \vphi + \de^i \vphi \, \de_j \vphi' \right]
      - \left( \1{\del} + 2 \vphi \right) \left( \de^i \vphi' + \H \de^i \vphi \right)
      \left( \de_j \vphi' + \H \de_j \vphi \right) \Big\} \, .
  \end{split}
\end{equation}

We now use the solutions at first and second order,
and the first order (0-0) equation $\1{\del} = - 2 \frac{g f}{\Om_m} \vphi_0$,
to obtain
\begin{equation}
  \begin{split}
    \nab^2 \3{\Q} =& \frac{1}{4 \pi G a^2 \rho_0}
    \left\{ E(\eta) \left[ 3 \nab^{-2} \de_i \de^j \left(\vphi_0 \, \de^i \vphi_0 \, \de_j \vphi_0 \right)
        - \left( \vphi_0 \, \de^k \vphi_0 \, \de_k \vphi_0 \right) \right] \right. \\
    & + \left. F(\eta) \left[ 3 \nab^{-2} \de_i \de^j
        \left( \de^i \al_0 \, \de_j \vphi_0 + \de^i \vphi_0 \de_j \al_0 \right)
       - 2 (\de^k \al_0 \, \de_k \vphi_0) \right] \right\} \, ,
  \end{split}
\end{equation}
where $E(\eta)$, $F(\eta)$, $\al_0(\x)$ are defined in eqs. \eqref{E}, \eqref{F}, \eqref{al0} respectively.

By differentiating the last equation we find also
\begin{equation}
  \begin{split}
    \nab^2 \3{\Q}' =& \H \nab^2 \3{\Q} + \frac{1}{4 \pi G a^2 \rho_0}
    \left\{ E' \left[ 3 \nab^{-2} \de_i \de^j \left(\vphi_0 \, \de^i \vphi_0 \, \de_j \vphi_0 \right)
        - \left( \vphi_0 \, \de^k \vphi_0 \, \de_k \vphi_0 \right) \right] \right. \\
    & + \left. F' \left[ 3 \nab^{-2} \de_i \de^j
        \left( \de^i \al_0 \, \de_j \vphi_0 + \de^i \vphi_0 \de_j \al_0 \right)
       - 2 (\de^k \al_0 \, \de_k \vphi_0) \right] \right\} \, ,
  \end{split}
\end{equation}
using the background equation $\rho_0' = - 3 \H \rho_0$;
and
\begin{equation}
  \begin{split}
    \3{T} =& \frac{1}{(4 \pi G a^2)^2 \rho_0}
    \Big[ \left( g' + \H g \right) \left(2 A' + \H A + \H C - 2 g g' + 2 \H \frac{g^2 f^2}{\Om_m} \right)
      \left( \vphi_0 \, \de^k \vphi_0 \, \de_k \vphi_0 \right) \\
      & + \left( g' + \H g \right) B' \, \left( \de^k \al_0 \, \de_k \vphi_0 \right) \Big] \, .
  \end{split}
\end{equation}
where $A(\eta)$, $B(\eta)$, $C(\eta)$ are defined in eqs. \eqref{A}, \eqref{B}, \eqref{C} respectively.

Finally, putting all these expressions together, we arrive at eq. \eqref{psi},
while eq. \eqref{relationphipsi} is simply given by expanding eq. \eqref{tracelessnp}.


\begin{references} 
\bibitem{kwmap3}
  D.~N.~Spergel {\it et al.}  [WMAP Collaboration],
  %``Wilkinson Microwave Anisotropy Probe (WMAP) three year results:
  %Implications for cosmology,''
  arXiv:astro-ph/0603449.
  %%CITATION = ASTRO-PH/0603449;%% 


\bibitem{Planck}
See http://planck.esa.int/.


\bibitem{review}
  N.~Bartolo, E.~Komatsu, S.~Matarrese and A.~Riotto,
  %``Non-Gaussianity from inflation: Theory and observations,''
  Phys.\ Rept.\  {\bf 402}, 103 (2004).

\bibitem{ABMR} 
V.~Acquaviva, N.~Bartolo, S.~Matarrese and A.~Riotto,
%``Second-order cosmological perturbations from inflation,''
Nucl.\ Phys.\ B {\bf 667}, 119 (2003). 
%%CITATION = ASTRO-PH 0209156;%%

\bibitem{maldacena}
J.~Maldacena,
%``Non-Gaussian features of primordial fluctuations in single field
%inflationary models,''
JHEP {\bf 0305}, 013 (2003).
%%CITATION = ASTRO-PH 0210603;%%




\bibitem{MGLM}
  S.~Mollerach, A.~Gangui, F.~Lucchin and S.~Matarrese,
  %``Contribution to the three point function of the cosmic microwave background
  %from the Rees-Sciama effect,''
  Astrophys.\ J.\  {\bf 453}, 1 (1995)
  [arXiv:astro-ph/9503115].
  %%CITATION = ASJOA,453,1;%%

\bibitem{Bern}
  F.~Bernardeau,
  %``Weak Lensing Detection in CMB Maps,''
  Astron.\ Astrophys.\  {\bf 324}, 15 (1997).

\bibitem{GS1}
  D.~N.~Spergel and D.~M.~Goldberg,
  %``Microwave background bispectrum. 1. Basic formalism,''
  Phys.\ Rev.\  D {\bf 59}, 103001 (1999).


\bibitem{GS2}
D.~M.~Goldberg and D.~N.~Spergel,
  %``Microwave background bispectrum. 2. A probe of the low redshift universe,''
  Phys.\ Rev.\  D {\bf 59}, 103002 (1999).




\bibitem{ks}
E.~Komatsu and D.~N.~Spergel,
%``Acoustic signatures in the primary microwave background bispectrum,''
Phys.\ Rev.\ D {\bf 63}, 063002 (2001).


\bibitem{VerdeSDE}
  L.~Verde and D.~N.~Spergel,
  %``Dark energy and cosmic microwave background bispectrum,''
  Phys.\ Rev.\  D {\bf 65}, 043007 (2002).


\bibitem{GBP}
  F.~Giovi, C.~Baccigalupi and F.~Perrotta,
  %``Cosmic microwave background constraints on dark energy dynamics: beyond the
  %power spectrum analysis,''
  Phys.\ Rev.\  D {\bf 71}, 103009 (2005).
 


\bibitem{fullT}
  N.~Bartolo, S.~Matarrese and A.~Riotto,
  %``The Full Second-Order Radiation Transfer Function for Large-Scale CMB
  %Anisotropies,''
  JCAP {\bf 0605}, 010 (2006).



\bibitem{kwmap}
E.~Komatsu {\it et al.},
%``First Year Wilkinson Microwave Anisotropy Probe (WMAP) Observations: Tests
%of Gaussianity,''
Astrophys.\ J.\ Suppl.\  {\bf 148}, 119 (2003).


\bibitem{babichpol}
  D.~Babich and M.~Zaldarriaga,
  %``Primordial Bispectrum Information from CMB Polarization,''
  Phys.\ Rev.\  D {\bf 70}, 083005 (2004).


\bibitem{Liguorietal}
  M.~Liguori, F.~K.~Hansen, E.~Komatsu, S.~Matarrese and A.~Riotto,
  %``Testing Primordial Non-Gaussianity in CMB Anisotropies,''
  Phys.\ Rev.\  D {\bf 73}, 043505 (2006).

\bibitem{Liguorietal2}
M.~Liguori, A.~Yadav, F.~K.~Hansen, E.~Komatsu, S.~Matarrese, B.~Wandelt., to appear. 

\bibitem{OkamotoHu}
  T.~Okamoto and W.~Hu,
  %``The Angular Trispectra of CMB Temperature and Polarization,''
  Phys.\ Rev.\  D {\bf 66}, 063008 (2002).






\bibitem{KogoKomatsu}
  N.~Kogo and E.~Komatsu,
  %``Angular Trispectrum of CMB Temperature Anisotropy from Primordial
  %Non-Gaussianity with the Full Radiation Transfer Function,''
  Phys.\ Rev.\  D {\bf 73}, 083007 (2006).


\bibitem{Bartolo:2005fp}
  N.~Bartolo, S.~Matarrese and A.~Riotto,
  %``Non-Gaussianity of Large-Scale CMB Anisotropies beyond Perturbation
  %Theory,''
  JCAP {\bf 0508} (2005) 010



\bibitem{SL1}
  D.~Seery, J.~E.~Lidsey and M.~S.~Sloth,
  %``The inflationary trispectrum,''
  JCAP {\bf 0701}, 027 (2007).




\bibitem{SL2}
  D.~Seery and J.~E.~Lidsey,
  %``Non-gaussianity from the inflationary trispectrum,''
  JCAP {\bf 0701}, 008 (2007).




\bibitem{SasakiVW}
  M.~Sasaki, J.~Valiviita and D.~Wands,
  %``Non-gaussianity of the primordial perturbation in the curvaton model,''
  Phys.\ Rev.\  D {\bf 74}, 103003 (2006).


\bibitem{BSW}
  C.~T.~Byrnes, M.~Sasaki and D.~Wands,
  %``The primordial trispectrum from inflation,''
  Phys.\ Rev.\  D {\bf 74}, 123519 (2006).








\bibitem{HuangShiu}
  M.~x.~Huang and G.~Shiu,
  %``The inflationary trispectrum for models with large non-Gaussianities,''
  Phys.\ Rev.\  D {\bf 74}, 121301 (2006).





\bibitem{Crem}
  P.~Creminelli, L.~Senatore, M.~Zaldarriaga and M.~Tegmark,
  %``Limits on f_NL parameters from WMAP 3yr data,''
  JCAP {\bf 0703}, 005 (2007).





\bibitem{Mollerach:1997up}
  S.~Mollerach and S.~Matarrese,
  %``Cosmic microwave background anisotropies from second order gravitational
  %perturbations,''
  Phys.\ Rev.\  D {\bf 56}  4494 (1997).



\bibitem{Pyne:1995bs}
  T.~Pyne and S.~M.~Carroll,
  %``Higher-Order Gravitational Perturbations of the Cosmic Microwave
  %Background,''
  Phys.\ Rev.\  D {\bf 53},  2920 (1996).
  %%CITATION = PHRVA,D53,2920;%%


\bibitem{Pyne:1993np}
  T.~Pyne and M.~Birkinshaw,
  %``Null geodesics in perturbed space-times,''
  arXiv:astro-ph/9303020.
  %%CITATION = ASTRO-PH/9303020;%%



  


\bibitem{CMB2first}
N.~Bartolo, S.~Matarrese and A.~Riotto,
  %``CMB Anisotropies at Second Order I,''
  JCAP {\bf 0606}, 024 (2006).




\bibitem{CMB2second}
  N.~Bartolo, S.~Matarrese and A.~Riotto,
  %``CMB Anisotropies at Second-Order II: Analytical Approach,''
  JCAP {\bf 0701}, 019 (2007).


\bibitem{MMB}
  S.~Matarrese, S.~Mollerach and M.~Bruni,
  %``Second-order perturbations of the Einstein-de Sitter universe,''
  Phys.\ Rev.\  D {\bf 58}, 043504 (1998).

\bibitem{numerico}
 N.~Bartolo, E.~Komatsu, S.~Matarrese, D~Nitta and A.~Riotto, to appear.

\bibitem{MHMpol}
  S.~Mollerach, D.~Harari and S.~Matarrese,
  %``CMB polarization from secondary vector and tensor modes,''
  Phys.\ Rev.\  D {\bf 69}, 063002 (2004).



\bibitem{SalopekBond}
D.~S.~Salopek, J.~R.~Bond, Pys.~Rev.~{\bf D} 42, 3936 (1990).


\bibitem{CZRec}
  P.~Creminelli and M.~Zaldarriaga,
  %``CMB 3-point functions generated by non-linearities at recombination,''
  Phys.\ Rev.\  D {\bf 70}, 083532 (2004).
 


\bibitem{lahav}
O.~Lahav, P.~B.~Lilje, J.~R.~Primack and M.~J.~Rees,
%``Dynamical effects of the cosmological constant,''
Mon.\ Not.\ Roy.\ Astron.\ Soc.\  {\bf 251}, 128 (1991).
%%CITATION = MNRAA,251,128;%%


\bibitem{Carroll}
S.~M.~Carroll, W.~H.~Press and E.~L.~Turner,
%``The Cosmological constant,''
Ann.\ Rev.\ Astron.\ Astrophys.\  {\bf 30}, 499 (1992).
%%CITATION = ARAAA,30,499;%%


\bibitem{Eisenstein}
D.~J.~Eisenstein,
%``An Analytic Expression for the Growth Function in a Flat Universe with a
%Cosmological Constant,''
arXiv:astro-ph/9709054.
%%CITATION = ASTRO-PH 9709054;%%




\bibitem{BMR2}
N.~Bartolo, S.~Matarrese and A.~Riotto,
%``Enhancement of non-Gaussianity after inflation,''
JHEP {\bf 0404}, 006 (2004). 
%%CITATION = ASTRO-PH 0308088;%%



\bibitem{lrreview} For a review, see
D.~H.~Lyth and A.~Riotto,
%``Particle physics models of inflation and the cosmological density
%perturbation,''
Phys.\ Rept.\  {\bf 314}, 1 (1999) 
%%CITATION = HEP-PH 9807278;%%

\bibitem{BST}
J.M. Bardeen, P. J. Steinhardt, M. S. Turner, Phys.~Rev. {\bf D28}, 679 (1983).




\bibitem{KMNR}
  E.~W.~Kolb, S.~Matarrese, A.~Notari and A.~Riotto,
  %``Cosmological influence of super-Hubble perturbations,''
  Mod.\ Phys.\ Lett.\  A {\bf 20}, 2705 (2005)




\bibitem{LLMR}
  J.~Lesgourgues, M.~Liguori, S.~Matarrese and A.~Riotto,
  %``CMB lensing extraction and primordial non-Gaussianity,''
  Phys.\ Rev.\  D {\bf 71}, 103514 (2005).
 

\bibitem{prl}
  N.~Bartolo, S.~Matarrese and A.~Riotto,
  %``Gauge-invariant temperature anisotropies and primordial  non-Gaussianity,''
  Phys.\ Rev.\ Lett.\  {\bf 93}, 231301 (2004).


\end{references}
\end{document}